\documentstyle[psfig,sprocl]{article}
\makeatletter
\newif\if@cghi
\def\@cite#1#2{{\if@cghi$\!^{#1}$\if@tempswa\typeout
        {IJCGA warning: optional citation argument 
        ignored: `#2'} \fi\else [{#1\if@tempswa , #2\fi}]\fi}}
\makeatother

\newcommand{\col}  {Collaboration}
\newcommand{\etal} {{\em et al.}}
\newcommand{\ibid} {{\em ibid.}}
\newcommand{\ie}   {{\em i.e.}}
\newcommand{\eg}   {{\em e.g.}}
\newcommand{\plb}[2]{{\em Phys. Lett.}          {\bf #1B}, #2 }
\newcommand{\npb}[2]{{\em Nucl. Phys.}          {\bf B#1}, #2 }
\newcommand{\pr }[2]{{\em Phys. Rep.}           {\bf  #1}, #2 }
\newcommand{\prd}[2]{{\em Phys. Rev.}           {\bf D#1}, #2 }
\newcommand{\ptp}[2]{{\em Prog. Theo. Phys.}    {\bf D#1}, #2 }
\newcommand{\prl}[2]{{\em Phys. Rev. Lett.}     {\bf  #1}, #2 }
\newcommand{\ncl}[2]{{\em Nuovo Cimento Lett.}  {\bf  #1}, #2 }
\newcommand{\zpc}[2]{{\em Z. Phys.}             {\bf C#1}, #2 }
\newcommand{\sci}[2]{{\em Science}              {\bf  #1}, #2 }
\newcommand{\epj}[2]{{\em Eur. Phys. J.}        {\bf  #1}, #2 }
\newcommand{\app}[2]{{\em Acta Phys. Polon.}    {\bf B#1}, #2 }
\newcommand{\mpl}[2]{{\em Mod. Phys. Lett.}     {\bf A#1}, #2 }
\newcommand{\cpc}[2]{{\em Comp. Phys. Comm.}	{\bf  #1}, #2 }
\newcommand{\con}[2]{                           {\bf  #1}, #2 }
\def\L{{\cal L}}
\def\nn{\nonumber}
\def\oa{${\cal O}(\alpha)$}
\def\ef{{\rm eff}}
\def\slash#1{#1\!\!\!{/}}
\renewcommand{\bar}[1]{\,\overline{#1}}
\def\fiv{${\bf 5}+{\bf\bar5}$}
\def\ten{${\bf 10}+{\bf\bar{10}}$}
\newcommand{\half}{\frac{1}{2}}
\newcommand{\gsim} {\buildrel > \over {_\sim}}

\def\re{{\cal R}e\,}
\newcommand{\dr}{\mbox{{\footnotesize{$\overline{{\rm DR}}$}}}}
\newcommand{\ms}{\mbox{{\footnotesize{$\overline{{\rm MS}}$}}}}

\begin{document}

\title{RENORMALIZATION OF SUPERSYMMETRIC THEORIES}

\author{DAMIEN M. PIERCE}

\address{Stanford Linear Accelerator Center\\ Stanford University\\
Stanford, CA 94309, USA\\ E-mail: pierce@slac.stanford.edu}

\maketitle

\abstracts{We review the renormalization of the electroweak sector of
the standard model. The derivation also applies to the minimal
supersymmetric standard model. We discuss regularization, and the
relation between the threshold corrections and the renormalization
group equations. We consider the corrections to many precision
observables, including $M_W$ and $\sin^2\theta^{\rm eff}$. We show
that global fits to the data exclude regions of supersymmetric model
parameter space and lead to lower bounds on superpartner masses.}

\section{Introduction}

\subsection{The standard model and beyond}

Over the past decade we have seen the standard model (SM) tested with
increasing precision by a large number of experiments. It is quite
remarkable that as the measurements become more accurate and encompass
more observables, the standard model is repeatedly confirmed to
greater precision. When discrepancies are found, they always seem to
evaporate with subsequent data and/or analyses. From the point of view
of precision measurements there is no need for physics beyond the
standard model.

With so much direct experimental evidence confirming the standard
model (and no evidence of discrepancies \cite{note}) it might appear
somewhat mysterious that so much time and effort is spent studying
models beyond the standard model. There are many reasons why people
believe the standard model with a light Higgs boson is not the whole
story. Many agree that new physics near the electroweak scale is
necessary in order to resolve the hierarchy problem. They point out
that in order for the huge hierarchy between the electroweak and
Planck scales to exist the parameters in the Higgs sector of the
standard model must be tuned to one part in $10^{34}$. New physics
near the electroweak scale can stabilize or obviate this hierarchy. As
the scale of new physics is raised above the TeV scale, either a new
(albeit much less severe) hierarchy problem arises or there are
problems with perturbative unitarity. Another major reason to consider
physics beyond the standard model is in order to find a simpler or
unified theory which offers an explanation for the standard model
symmetries, matter content, and/or gives some framework for
understanding the values of the 18 standard model input parameters.

Models of physics beyond the standard model can for the most part be
divided into two classes, which are distinguished by their
implementation of electroweak symmetry breaking. These are technicolor
theories and supersymmetry. The simplest (and, it turns out, unviable)
technicolor models operate in many respects like a scaled-up version
of QCD. Such models typically encounter problems related to the
fermion mass hierarchy, flavor changing neutral currents, electroweak
precision tests, and/or light pseudoscalars. To avoid this plethora of
pitfalls, rather large, complicated, and seemingly {\it ad hoc} models
must be invoked.

Supersymmetry, on the other hand, does not suffer from any of the
previously mentioned problems. In fact, due to the decoupling nature
of supersymmetric corrections, the minimal supersymmetric standard
model (MSSM) with a heavy ($\sim$ 1 TeV) superpartner spectrum is
indistinguishable from the standard model in all weak-scale
experiments. In this light, the success of the standard model can be
also realized as the success of the MSSM.

The fact that the gauge couplings nearly unify in the MSSM \cite{gcu}
(but not the standard model) may be taken as indirect evidence for
supersymmetric grand unified theories (GUTs). String theory is the
only viable candidate for a theory of quantum gravity, and
supersymmetry remains (for the most part) an essential element in the
construction of sensible string theories.  In some string models,
low-energy supersymmetry is an unavoidable by-product.  Gauge coupling
unification can arise in string theory, with or without a GUT.

Another nice virtue of supersymmetry is related to electroweak
symmetry breaking. In the standard model, electroweak symmetry
breaking is accomplished by setting the Higgs mass parameter to the
wrong-sign by hand. In popular supersymmetric models, however, the
Higgs mass parameter has a positive value at an initial (high)
scale. By virtue of radiative corrections the mass runs with the
scale, and at the electroweak scale it has the wrong sign. This
radiative breaking of electroweak symmetry is a rather generic feature
of supersymmetric models.

While supersymmetric theories readily explain electroweak symmetry
breaking, they introduce the problem of supersymmetry breaking. {\it A
priori} there are hundreds of arbitrary new parameters in the
supersymmetry-breaking Lagrangian. There are simple models of
supersymmetry breaking which greatly reduce the number of
parameters. For example, in the simplest gravity- and gauge-mediated
models that we will consider, there are about 5 input parameters.

With this introduction, we are motivated to study in some detail the
ramifications of low-energy supersymmetry. Radiative corrections play
an essential role in much of the discussion of supersymmetry
phenomenology. In the next section we will pedagogically review
renormalization of the electroweak sector of the standard model.  This
discussion applies to the MSSM as well. In the following section we
discuss regularization, threshold corrections, and the renormalization
group equations (RGEs). We apply these results when we discuss the
global fit of electroweak data in the standard model and the MSSM in
Sec.~4. A brief summary and conclusions are provided in Sec.~5.

\section{Electroweak Renormalization}
\label{sec.ren}

Renormalization remains a cornerstone of particle physics. In any
model of particle physics, whether the standard model or beyond, if
radiative corrections are to be taken into account some
renormalization procedure must be implemented. Here we review the
standard technique of counterterm renormalization.  We will focus on
the renormalization necessary to calculate the supersymmetric
corrections to electroweak observables. For this we need not consider
full gauge-sector renormalization (ghosts; gauge-fixing). For a more
complete treatment see Ref.~\citelow{renormalization}.

We first introduce the counterterms, and then discuss the
renormalization conditions which determine them. In the following
section we consider the calculation of the counterterms (and Feynman
diagrams in general) by discussing regularization and
Passarino-Veltman functions.

The renormalization procedure is straightforward. We start in each
case with the bare Lagrangian, which consists of bare parameters (bare
masses and couplings) and bare fields. We write this in terms of the
renormalized parameters and fields. For example, the bare mass $m_b$
and coupling $e_b$ are replaced by the renormalized parameters $m$ and
$e$, and the associated counterterms, $\delta m$ and $\delta e$,
$$m_b = m + \delta m\ ,\qquad e_b = e + \delta e~.$$ The bare fields
are equal to the renormalized fields multiplied by a wave-function
renormalization factor, \eg\
$$\phi_b = Z^\half\phi~.$$ The $Z^\half$ factor is in general a
matrix, mixing fields with the same quantum numbers into each other.
After these replacements, we can separate the bare Lagrangian into the
renormalized Lagrangian and the counterterm Lagrangian. The
counterterm Lagrangian gives rise to new vertices, which lead to new
graphs in each order of perturbation theory. These counterterm graphs
cancel subdivergences in multiloop diagrams. At one-loop order,
however, the counterterm vertices enter only in tree graphs. Hence,
the one-loop renormalization procedure is quite simple, and amounts to
adding the one-loop diagram contribution to the tree-level (bare)
Lagrangian contribution. In the last step of the renormalization
procedure we apply the renormalization conditions. The renormalization
conditions determine the counterterms and the physical meaning of the
renormalized parameters.

For mass and wave-function renormalization, it is most natural to
impose the on-shell renormalization conditions. In this case the
renormalized mass is the pole mass (\ie\ the experimentally
measured mass). The on-shell renormalization conditions consist of two
parts:
\begin{eqnarray}
&&\hspace{-.3in}\mbox{
a) the renormalized mass is the real part of the pole of the
propagator,}\label{a}\\
&&\hspace{-.3in}\mbox{
b) the real part of the residue of the pole is unity.}\label{b}
\end{eqnarray}
In the general case the inverse propagator is a matrix. The bosonic
form is
\begin{equation} D^{-1}(p^2) = \left(\begin{array}{rrc}
p^2-m^2_{11}+\Pi_{11}(p^2)\quad & - m^2_{12} + \Pi_{12}(p^2)\quad &
\cdots\\ -m^2_{21}+\Pi_{21}(p^2)\quad &
p^2-m^2_{22}+\Pi_{22}(p^2)\quad & \cdots \\ \vdots\qquad\qquad &
\vdots\qquad\qquad & \cdots \end{array}\right)\ .
\end{equation} Here $p^2$ is the external momenta-squared, $m_{ij}$ is
the tree-level (bare) mass matrix and $\Pi_{ij}(p^2)$ is the one-loop
self-energy. The poles $p_i^2$ are determined from $${\rm
Det}\left(D^{-1}(p^2)\right)=0\ ,$$ and the pole masses are $m_i^2 =
\re(p_i^2)$.

Scalar mass and wave-function renormalization provides the simplest
example. The quadratic part of the bare Lagrangian is
\begin{eqnarray} 
\L &=& \partial_\mu\phi^*_b\partial^\mu\phi_b - m^2_b\phi^*_b\phi_b\nn\\
   &=& Z_\phi\partial_\mu\phi^*\partial^\mu\phi -
       Z_\phi(m^2+\delta m^2)\phi^*\phi~.\nn\\
\end{eqnarray}
So, the bare inverse propagator can be read off, and at one-loop level
we simply add the self-energy to obtain the renormalized inverse
propagator,
$$D^{-1}(p^2) = Z_\phi p^2 - Z_\phi(m^2 + \delta m^2) + \Pi(p^2)\ .$$
The real part of the pole of the propagator is
$$
\re(p^2) = m^2 + \delta m^2 - Z_\phi^{-1}\re\Pi(p^2)~.
$$
The on-shell renormalization condition requires $\re(p^2)=m^2$. This
implies
\begin{equation}
\delta m^2 = Z_\phi^{-1}\re\Pi(p^2)~.
\end{equation}
At \oa\ we can set $Z_\phi^{-1}=1$ and $p^2=m^2$ in this equation, to
obtain
\begin{equation}
\delta m^2 = \re\Pi(m^2)~.
\end{equation}
The residue of the pole of the propagator is
\begin{equation}
{\rm res}\left(D(p^2)\right) = Z_\phi + \Pi'(p^2)~.
\end{equation}
The prime denotes the derivative with respect to $p^2$.  The second
part of the on-shell renormalization conditions requires that the real
part of the residue be equal to unity,
\begin{equation}
Z_\phi +\re\Pi'(p^2) = 1~.
\end{equation}
Again, at \oa\ we can replace the complex pole $p^2$ by the real part,
the renormalized mass, so that
\begin{equation}
Z_\phi = 1 - \re\Pi'(m^2)~.
\end{equation}

Next we consider gauge-boson mass and wave-function renormalization.
In this case the inverse propagator has a longitudinal and transverse
part,
\begin{equation}
D_{\mu\nu}(p^2) = \left(-g_{\mu\nu}+\frac{p_\mu p_\nu}{p^2}\right)
\ D^T(p^2) + \frac{p_\mu p_\nu}{p^2}\ D^L(p^2)\ .
\end{equation}
Only the transverse part of the propagator contributes to physical
processes. Correspondingly, only the transverse part of the one-loop
self energy will appear in the corrections.

The gauge boson renormalization is quite similar to the scalar case we
just considered. In the standard model $\gamma$-$Z$ mixing introduces
a slight complication. The quadratic part of the bare Lagrangian
involving $\gamma$ and $Z$ is
\begin{equation}
{\cal L} = (A^\mu_b\quad Z^\mu_b)
\left(
\begin{array}{cc}
\Box & 0 \\
\quad0\quad & \Box-M^2_{Z_b}
\end{array}
\right)
\left({A_{\mu_b}\atop Z_{\mu_b}}\right)~,
\end{equation}
where $\Box$ denotes the D'Alembertian.  As usual we rewrite this in
terms of the renormalized parameters and fields,
\begin{eqnarray}
&&{\cal L} =\nn\\
&&\left(A^\mu~~Z^\mu\right) \left(\begin{array}{cc}
Z_{\gamma\gamma}^\half & Z_{Z\gamma}^\half \\ Z_{\gamma Z}^\half &
Z_{ZZ}^\half \end{array}\right) 
\left(\begin{array}{cc} \Box & 0 \\ \qquad0\qquad & \Box-M^2_Z-\delta
M_Z^2\end{array}\right)
\left(\begin{array}{cc}
Z_{\gamma\gamma}^\half & Z_{\gamma Z}^\half \\ Z_{Z \gamma}^\half &
Z_{ZZ}^\half \end{array}\right) \left({A^\mu\atop Z^\mu}\right)\nn\\
&&= \left(A^\mu Z^\mu\right) \left(\begin{array}{cc}
Z_{\gamma\gamma}\Box & Z_{\gamma Z}^\half\Box +
Z_{Z\gamma}^\half(\Box-M_Z^2-\delta M_Z^2)\\
Z_{\gamma Z}^\half\Box + Z_{Z\gamma}^\half(\Box-M_Z^2-\delta M_Z^2) 
& Z_{ZZ}(\Box-M_Z^2-\delta M_Z^2)\end{array}\right)\nn\\
&&\hspace{3.8in}\times\left({A^\mu\atop Z^\mu}\right)~.\nn
\end{eqnarray}
We have dropped some terms of ${\cal O}(\alpha^2)$ such as
$(1-Z_{\gamma\gamma}^\half)Z_{\gamma Z}^\half$. The inverse propagator
may be read off as
\begin{equation}
D^{-1}(p^2) = \left(\begin{array}{cc} Z_{\gamma\gamma}p^2 +
\Pi_{\gamma\gamma}^T(p^2) & D^{-1}_{\gamma Z}(p^2)\\
D^{-1}_{Z\gamma}(p^2) & Z_{ZZ}(p^2 - M_Z^2 - \delta M_Z^2) +
\Pi_{ZZ}^T(p^2)\end{array}\right) \ ,
\end{equation}
where
\begin{equation}
D^{-1}_{\gamma Z}(p^2) = D^{-1}_{Z\gamma}(p^2) = 
Z_{\gamma Z}^\half p^2 + Z_{Z\gamma}^\half
(p^2-M_Z^2-\delta M_Z^2) + \Pi_{\gamma Z}^T(p^2)~,
\end{equation}
and we added the transverse part of the gauge-boson self-energy,
$\Pi^T(p^2)$.

We determine the counterterms by applying the on-shell renormalization
conditions, (\ref{a}--\ref{b}), which in this case can be realized
by$\,$\footnote{We use the fact that a self-energy evaluated at zero
external momentum is real.}
\begin{eqnarray}
&\mbox{a)\ }& D_{\gamma\gamma}^{-1}(0)=0\\ 
&& \re D_{ZZ}^{-1}(M_Z^2) = 0\\
&& D_{\gamma Z}^{-1}(0) = \re D_{\gamma Z}^{-1}(M_Z^2) = 0\\
&\mbox{b)\ }& \re {D_{\gamma\gamma}^{-1}}'(0) = \re {D_{ZZ}^{-1}}'(M_Z^2) = 1
\end{eqnarray}
These determine the counterterms
\begin{eqnarray}
\delta M_Z^2 &=& \re\Pi_{ZZ}^T(M_Z^2)\ ,\label{dmz}\\
Z_{\gamma\gamma} &=& 1 - \re{\Pi_{\gamma\gamma}^{T'}}(0)\ ,\\ 
Z_{ZZ} &=& 1 - \re{\Pi_{ZZ}^{T'}}(M_Z^2)\ ,\\
Z_{\gamma Z}^{\half} &=& -\re{\Pi_{\gamma Z}^T(M_Z^2)\over M_Z^2}~,\\
Z_{Z\gamma}^{\half} &=& {\Pi^T_{\gamma Z}(0)\over M_Z^2}~.\label{zg}
\end{eqnarray}
Note that no photon mass counterterm is required since the U(1) Ward
identity ensures $\Pi_{\gamma\gamma}^T(0)=0$. For the $W$-boson we
find
\begin{eqnarray}
\delta M_W^2  &=& \re\Pi_{WW}^T(M_W^2)\ ,\\ 
Z_{WW} &=& 1 - \re{\Pi_{WW}^{T'}}(M_W^2)\ .
\end{eqnarray}

The final mass and wave-function renormalization we need to consider
applies to fermion fields. Just as in the previous examples, we start
by writing the bare Lagrangian, then substitute for the renormalized
fields and mass. In general we will have mixing among the fields, so
the masses and the wave-function renormalization constants will be
matrices. The bare kinetic Lagrangian in the mass basis
is$\,$\footnote{Note in the mass basis $m$ and $\delta m$ are real and
diagonal.}
\begin{eqnarray}
\L_{\rm bare} &=& i\bar{\psi}_{b_L}\slash\partial
\psi_{b_L} + i\bar{\psi}_{b_R} \slash\partial\psi_{b_R} -
\bar{\psi}_{b_L} m_b\psi_{b_R} - \bar{\psi}_{b_R} m_b\psi_{b_L}
\nn\\[1ex]
&=& i\bar\psi_L Z^{\half\dagger}_L\slash\partial Z^{\half}_L\psi_L +
i\bar\psi_R Z^{\half\dagger}_R\slash\partial Z^{\half}_R \psi_R \nn \\[1ex] 
&& - \bar \psi_L Z^{\half\dagger}_L(m+\delta m)Z^{\half}_R\psi_R - \bar
\psi_R Z^{\half\dagger}_R (m+\delta m)Z_L\psi_L~,\nn
\end{eqnarray}
where $\psi_{L,R} = (1\mp\gamma_5)/2~\psi$.  From this expression we
can read off the bare inverse propagator, and at one-loop we simply
add the self-energy contribution ($\Sigma_L,\ \Sigma_R,\ \Sigma_S$) to
obtain the renormalized inverse propagator$\,$\footnote{We assumed
hermiticity, which implies that ${\Sigma_L}_{ii}$ and
${\Sigma_R}_{ii}$ are real and ${\Sigma_S}_R =
{\Sigma_S}_L^\dagger$. If the fermion under consideration is unstable
the effective Lagrangian is not Hermitian.},
\begin{eqnarray}
S^{-1}_{ij}(\slash p) &=&
\left(Z_L^{\half\dagger}Z_L^{\half}\right)_{ij} \slash p P_L +
\left(Z_R^{\half\dagger}Z_R^{\half}\right)_{ij}\slash p P_R\nn\\ &&-\
(Z^{\half\dagger}_{R\,ik} m_k Z^{\half}_{L\,kj} + \delta
m_i\,\delta_{ij})P_L\nn\\ &&-\
(Z^{\half\dagger}_{L\,ik}m_kZ^{\half}_{R\,kj}+\delta
m_i\,\delta_{ij})P_R\nn\\ &+&\Sigma_{L\,ij}(p^2)\slash p P_L +
\Sigma_{R\,ij}(p^2)\slash p P_R + \Sigma_{S\,ij}(p^2)P_L +
\Sigma^*_{S\,ji}(p^2)P_R~.\nn
\end{eqnarray}
We can define $Z_{L,R}^{\half}\equiv 1+\delta Z_{L,R}/2$. Then
we have
\begin{eqnarray}
S^{-1}_{ij}(\slash p) &=& 
\biggl[\delta_{ij} + \half\left(
\delta Z_L + \delta Z_L^\dagger\right)_{ij} +
\Sigma_{L\,ij}(p^2)\biggr]\slash p P_L \nn\\
&+&
\biggl[\delta_{ij} + \half\left(
\delta Z_R + \delta Z_R^\dagger\right)_{ij} +
\Sigma_{R\,ij}(p^2)\biggr]\slash p P_R \nn\\
&-& \biggl[m_i\delta_{ij} + 
\half\delta Z_{R\,ij}^\dagger m_j + 
\half m_i\delta Z_{L\,ij} + \delta m_i\,\delta_{ij}
- \Sigma_{S\,ij}(p^2)\biggr]P_L\nn\\
&-& \biggl[m_i\delta_{ij} + 
\half\delta Z_{L\,ij}^\dagger m_j + 
\half m_i\delta Z_{R\,ij} + \delta m_i\,\delta_{ij}
- \Sigma^*_{S\,ji}(p^2)\biggr]P_R~.\nn
\end{eqnarray}

To apply the on-shell renormalization conditions it is convenient to
introduce an on-shell spinor $u$ which satisfies the equation of
motion $(\slash p-m)u=0$.  Recall that the on-shell renormalization
conditions have two parts. The first part requires that the poles of
the renormalized propagator are at the renormalized masses. A pole in
the propagator corresponds to a vanishing column in the inverse
propagator, when the spinor multiplies the inverse propagator on the
right. We implement this condition by requiring
\begin{equation}
S_{ij}^{-1}(\slash p)u(m_j) = 0\label{fc1}
\end{equation}
for all $i,j$. The second part of the on-shell renormalization
conditions requires that the residues of the propagator poles are
unity. Note that Eq.~(\ref{fc1}) only requires that the diagonal
elements of the inverse propagator have the form $A(\slash p-m_i)$. To
ensure $A=1$, we require
\begin{equation}
{1\over\slash p-m_i}S_{ii}^{-1}(\slash p)u(m_i) = u(m_i)~.\label{fc2}
\end{equation}

From Eq.~(\ref{fc1}), with $i=j$, we have
\begin{eqnarray}
\half\left(\delta Z_L-\delta Z_R\right)_{ii}m_i-\delta m_i = 
-m_i\Sigma_{L\,ii}(m_i^2) - \Sigma^*_{S\,ii}(m_i^2)~,\\
\half\left(\delta Z_R-\delta Z_L\right)_{ii}m_i-\delta m_i = 
-m_i\Sigma_{R\,ii}(m_i^2) - \Sigma_{S\,ii}(m_i^2)~.
\end{eqnarray}
The sum and difference give
\begin{eqnarray}
\delta m_i &=& \half\biggl[\Sigma_{L\,ii}(m_i^2) + \Sigma_{R\,ii}(m_i^2)
\biggr]m_i + \re\Sigma_{S\,ii}(m_i^2)~,\label{dm}\\
\delta Z_{L\,ii} - \delta Z_{R\,ii}
&=& \Sigma_{R\,ii}(m_i^2)-\Sigma_{L\,ii}(m_i^2)
+{2\over m_i}\:{\cal I}m\Sigma_{S\,ii}(m_i^2)~.\label{dzlr}
\end{eqnarray}
With $i\ne j$, we find from Eq.~(\ref{fc1})
\begin{eqnarray}
\half\delta Z_{L\,ij}\,m_j - \half m_i\,\delta Z_{R\,ij}
+ \Sigma_{L\,ij}(m_j^2)\,m_j + \Sigma_{S\,ji}^*(m_j^2) &=& 0~,\\
\half\delta Z_{R\,ij}\,m_j - \half m_i\,\delta Z_{L\,ij}
+ \Sigma_{R\,ij}(m_j^2)\,m_j + \Sigma_{S\,ij}(m_j^2) &=& 0~.
\end{eqnarray}
This then determines
\begin{eqnarray}
&&\delta Z_{L\,ij} = \label{dzod1}\\
&&{2\over m_i^2-m_j^2}\biggl[
m_j^2\Sigma_{L\,ij}(m_j^2) + m_im_j\Sigma_{R\,ij}(m_j^2)
+ m_i\Sigma_{S\,ij}(m_j^2) + m_j\Sigma_{S\,ji}^*(m_j^2)\biggr]~,\nn\\
&&\delta Z_{R\,ij} =\label{dzod2}\\
&&{2\over m_i^2-m_j^2}\biggl[
m_j^2\Sigma_{R\,ij}(m_j^2) + m_im_j\Sigma_{L\,ij}(m_j^2)
+ m_j\Sigma_{S\,ij}(m_j^2) + m_i\Sigma_{S\,ji}^*(m_j^2)\biggr]~.\nn
\end{eqnarray}
Eq.~(\ref{fc2}) implies
\begin{eqnarray}
&&{1\over\slash p-m_i}\biggl[
\half\left(\delta Z_L + \delta Z_L^\dagger\right)_{ii}\slash p
- \half\left(\delta Z_L^\dagger + \delta Z_R\right)_{ii}m_i\\
&&\qquad\qquad-\delta m_i + \Sigma_{L\,ii}(p^2)\slash p 
+ \Sigma_{S\,ii}^*(p^2)\biggr]u(m_i)=0~,\nn\\
&&{1\over \slash p-m_i}\biggl[
\half\left(\delta Z_R + \delta Z_R^\dagger\right)_{ii}\slash p
- \half\left(\delta Z_R^\dagger + \delta Z_L\right)_{ii}m_i\\
&&\qquad\qquad-\delta m_i + \Sigma_{R\,ii}(p^2)\slash p 
+ \Sigma_{S\,ii}(p^2)\biggr]u(m_i)=0~.\nn
\end{eqnarray}
Taking the sum of these equations, and substituting $\delta m_i$
from Eq.~(\ref{dm}), we find
\begin{eqnarray}
&&{1\over\slash p-m_i}\biggl[
\half\left(\delta Z_L+\delta Z_L^\dagger + \delta Z_R+\delta
Z_R^\dagger\right)_{ii}(\slash p-m_i)\label{zlr}\\
&&\qquad+ \biggl(f(p^2)\slash p - f(m_i^2)m_i\biggr)
+ \biggl(g(p^2) - g(m_i^2)\biggr)
\biggr]u(m_i)=0\nn
\end{eqnarray}
where
\begin{equation}
f(p^2) = \Sigma_{L\,ii}(p^2)+\Sigma_{R\,ii}(p^2)~,\nn
\end{equation}
\begin{equation}
g(p^2) = 2\;\re\Sigma_{S\,ii}(p^2)~.\nn
\end{equation}
Note that
\begin{eqnarray}
{1\over\slash p-m_i}\biggl(f(p^2)\slash p-f(m_i^2)m_i\biggr)u(m_i)
&=& \biggl(f(m_i^2)+ 2m_i^2f'(m_i^2)\biggr)u(m_i)~,\nn\\
{1\over\slash p-m_i}\biggl(g(p^2)-g(m_i^2)\biggr)u(m_i)
&=& 2m_ig'(m_i^2)u(m_i)~,\nn
\end{eqnarray}
where $f'(m^2) = \partial f(p^2)/\partial p^2|_{p^2=m^2}$.
Eq.~(\ref{dzlr}) implies
\begin{equation}
\biggl(\delta Z_R+\delta Z_R^\dagger\biggr)_{ii} =
\biggl(\delta Z_L+\delta Z_L^\dagger\biggr)_{ii}
+2\biggl(\Sigma_L(m_i^2)-\Sigma_R(m_i^2)\biggr)_{ii}~.
\end{equation}
Substituting this in Eq.~(\ref{zlr}), we find
\begin{eqnarray}
\re\,\delta Z_{L\,ii} &=& -\Sigma_{L\,ii}(m_i^2) - m_i^2\left(
\Sigma_{L\,ii}'(m_i^2)+\Sigma_{R\,ii}'(m_i^2)\right)
- 2m_i\,\re\Sigma_{S\,ii}'(m_i^2)~,\nn\\
\re\,\delta Z_{R\,ii} &=& -\Sigma_{R\,ii}(m_i^2) - m_i^2\left(
\Sigma_{L\,ii}'(m_i^2)+\Sigma_{R\,ii}'(m_i^2)\right)
- 2m_i\,\re\Sigma_{S\,ii}'(m_i^2)~.\nn
\end{eqnarray}

Note that the Lagrangian is invariant if we rotate both the left and
right-handed fields $\psi_{L\,i},\ \psi_{R\,i}$ by phases
$e^{i\theta_i}$. This means that there is an arbitrariness in the
diagonal wave-function renormalization constants. The on-shell
renormalization conditions are still satisfied if we transform the
$\delta Z_{ii}$'s as
\begin{equation}
\delta Z_{L\,ii}\rightarrow e^{i\theta_i}\delta Z_{L\,ii}~,\qquad\qquad
\delta Z_{R\,ii}\rightarrow e^{i\theta_i}\delta Z_{R\,ii}~.
\end{equation}
No physical results depend on this arbitrariness. This freedom allows
us to choose the $\delta Z_{R\,ii}$ to be real. We then determine the
imaginary part of $\delta Z_{L\,ii}$ from Eq.~(\ref{dzlr}). We add it
to the real part to obtain
\begin{eqnarray}
\delta Z_{L\,ii} &=& -\Sigma_{L\,ii}(m_i^2) - m_i^2\left(
\Sigma_{L\,ii}'(m_i^2)+\Sigma_{R\,ii}'(m_i^2)\right)\\
&&\quad- 2m_i\,\re\Sigma_{S\,ii}'(m_i^2) + {2\over m_i}\,{\cal I}m
\Sigma_{S\,ii}(m_i^2)~.\nn
\end{eqnarray}

The fermion renormalization constants are somewhat complicated.  If
one introduces the effective mixing matrices, which diagonalize the
renormalized mass matrix, the wave-function renormalization takes a
much simpler form. In particular, the off-diagonal wave-function
renormalization constants (Eqs.~(\ref{dzod1}--\ref{dzod2})) are
absorbed in the construction of the effective mixing matrices. See
Ref.~\citelow{knpy} for details.

We will be primarily interested in the $e$ and $\mu$ lepton
wave-function renormalization.  In the standard model, and in the
supersymmetric extensions we will consider, ${\cal
I}m\,\Sigma_{S\,ii}=0$. Also, there is no lepton mixing, and we
can approximate $m_\ell=0$. In this case the lepton wave-function
renormalization takes the particularly simple form
\begin{equation}
\delta Z_{L} = -\Sigma_{L}(0)~,\qquad\qquad
\delta Z_{R} = -\Sigma_{R}(0)~.\label{fwfr}
\end{equation}
We will use Eq.~(\ref{fwfr}) repeatedly.

It will be convenient to write the wave-function renormalization in the
vector/axial-vector basis. Consider the wave-function renormalization of
$$\bar\psi_b\gamma_\mu(g_LP_L+g_RP_R)\psi_b =
\bar\psi\gamma_\mu(Z_Lg_LP_L+Z_Rg_RP_R)\psi~.$$ It is a simple
exercise to show that
$$\bar\psi\gamma_\mu(Z_Lg_LP_L+Z_Rg_RP_R)\psi =
\bar\psi\biggl[(g_vZ_v+g_aZ_a)-(g_vZ_a+g_aZ_v)\gamma_5\biggr]\psi~,$$
where
\begin{eqnarray}
&&g_v=\half(g_L+g_R)~, \qquad g_a=\half(g_L-g_R)~,\\
&&Z_v=\half(Z_L+Z_R)~, \qquad Z_a=\half(Z_L-Z_R)~.
\end{eqnarray}
Note that $Z_v=1+{\cal O}(\alpha)$ and $Z_a={\cal O}(\alpha)$.  Hence,
we introduce $\delta Z_v\equiv 1-Z_v$ and $\delta Z_a\equiv Z_a$.

With these gauge-boson and fermion mass and wave-function renormalizations,
it is straightforward to renormalize the electric charge. The bare Lagrangian
contains the interaction terms
\begin{eqnarray}
{\cal L}_{\rm int}=&&-e_bQ\bar\psi_b\gamma_\mu\psi_b A^\mu_b -
\bar\psi_b\gamma_\mu({g_v}_b-{g_a}_b\gamma_5)\psi_b Z^\mu_b\\ &=&
-(e+\delta e)\,Q\,\bar\psi\gamma_\mu(Z_v-Z_a\gamma_5)\psi
\left(Z_{\gamma\gamma}^\half A_\mu + Z_{\gamma Z}^\half Z^\mu\right)\\
&&-\bar\psi\gamma_\mu\biggl[(g_vZ_v+g_aZ_a)-(g_vZ_a+g_aZ_v)\gamma_5\biggr]
\psi\,\biggl(Z_{ZZ}^\half Z_\mu + Z_{Z\gamma}^\half A^\mu\biggr)~.\nn
\end{eqnarray}
The terms involving $A^\mu$ are (dropping terms of ${\cal
O}(\alpha^2)$),
\begin{eqnarray}
&&-e\,A^\mu\label{tvert}\\
&&\bar\psi\,\gamma_\mu\Biggl\{\biggl[Q\left(1+{\delta e\over
e}+\delta Z_v+\half\delta Z_{\gamma\gamma}\right) + {g_v\over
e}Z_{Z\gamma}^\half\biggr] - \biggl[Q\,\delta Z_a + {g_a\over
e}Z_{Z\gamma}^\half\biggr]\gamma_5\Biggr\}\psi~.\nn
\end{eqnarray}
There are three form factors in the proper fermion-photon vertex,
\begin{eqnarray}
&&i\Lambda_\mu(k^2,p_1^2,p_2^2) =\\
\quad&&-ie\biggl[V_1(k^2,p_1^2,p_2^2)\gamma_\mu -
V_2(k^2,p_1^2,p_2^2)\gamma_\mu\gamma_5 +
{i\sigma_{\mu\nu}k^\nu\over2m_e}V_3(k^2,p_1^2,p_2^2)\biggr]~.\nn
\end{eqnarray}
We can read off the tree-level vertex from Eq.~(\ref{tvert}) and at
one-loop we simply add the proper vertex to obtain the renormalized
vertex
\begin{eqnarray}
&&\Gamma_\mu(k^2,p_1^2,p_2^2)=\label{renv}\\
\quad&& -e\Biggl\{\gamma_\mu\biggl[Q\left(1+{\delta
e\over e}+\delta Z_v + \half\delta Z_{\gamma\gamma}\right) +{g_v\over
e}Z_{Z\gamma}^\half + V_1(k^2,p_1^2,p_2^2)\biggr]\nn\\
\quad&& -\gamma_\mu\gamma_5\biggl[Q\,\delta Z_a + {g_a\over
e}Z_{Z\gamma}^\half +V_2(k^2,p_1^2,p_2^2)\biggr] \ +\
{i\sigma_{\mu\nu}k^\nu\over2m_e}V_3(k^2,p_1^2,p_2^2)\Biggr\}~.\nn
\end{eqnarray}
The on-shell renormalization condition for the vertex requires
\begin{equation}
\Gamma_\mu(0,m_e^2,m_e^2) = -eQ\gamma_\mu -
{ie\sigma_{\mu\nu}k^\nu\over2m_e}V_3(0,m_e^2,m_e^2)~,\label{rene}
\end{equation}
with the renormalized charge $e$ equal to that measured in Thompson
scattering, $e^2/4\pi=1/137.065$. The absence of an axial-vector
contribution in Eq.~(\ref{rene}) is an automatic consequence of the
axial-vector Ward identity,
\begin{equation}
e\;Q\ \delta Z_a + e\,V_2(0,m_e^2,m_e^2) = -g_a\,Z_{Z\gamma}^\half~.
\end{equation}
There is a corresponding vector Ward identity,
\begin{equation}
e\;Q\ \delta Z_v + e\,V_1(0,m_e^2,m_e^2) 
= -g_a\,Z_{Z\gamma}^\half~.\label{vward}
\end{equation}
Diagramatically, these equations correspond to (ignoring the magnetic
moment contribution)

\vspace{.2in}\hspace{.02in}
\psfig{figure=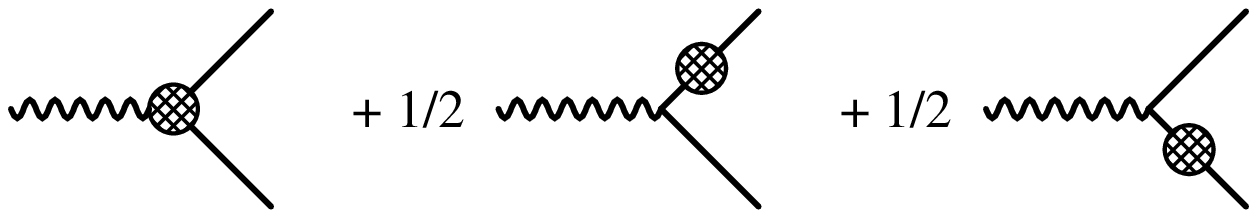,width=2.7in}

\vspace{-.35in}\hspace{2.96in}
$=~-g_a\,Z_{Z\gamma}^\half\gamma_\mu\,(1-\gamma_5)$
\vspace{.28in}

\noindent In QED the right hand side vanishes ($Z_{Z\gamma}^\half=0$)
and the usual QED Ward identity is realized. In SU(2)$\times$U(1), one
can consider the various contributions to the right hand side from the
various particles in the loop. From Eq.~(\ref{zg}),
$Z_{Z\gamma}^\half=\Pi^T_{\gamma Z}(0)/M_Z^2$.  At one-loop level, all
the contributions to $\Pi^T_{\gamma Z}(0)$ vanish save one, that of
the $W$-boson.  This is true in any model with the standard model
gauge symmetry. Hence, in the MSSM $\Pi^T_{\gamma Z}(0)$ has the same
value as in the standard model.

If we substitute Eq.~(\ref{vward}) into Eq.~(\ref{renv}), we can write
the on-shell vector coupling as
\begin{equation}
-eQ\gamma_\mu\biggl[1 + {\delta e\over e} + \half\delta
Z_{\gamma\gamma} + {g_v-g_a\over Qe}Z_{Z\gamma}^\half\biggr]
\end{equation}
which is equal to $-eQ\gamma_\mu$ by the renormalization condition
Eq.~(\ref{rene}). Hence,
\begin{eqnarray}
{\delta e\over e} &=& -\half\delta Z_{\gamma\gamma} - {g_v-g_a\over
Qe}Z_{Z\gamma}^\half\\
&=&\half\Pi^{T\prime}_{\gamma\gamma}(0) + {s\over c}{\Pi^T_{\gamma Z}(0)
\over M_Z^2}~,\label{de}
\end{eqnarray}
where $s$ ($c$) is the $\sin$ $(\cos)$ of the weak mixing angle.  It
is customary to define $\Delta\alpha = \delta\alpha/\alpha =
2\delta e/e$.

Note that the vector Ward identity, Eq.~(\ref{vward}), ensures that
the renormalization of the electric charge is universal, \ie\ the
flavor dependent wave-function and vertex corrections drop out.

Building on these results, we can readily derive the relationships
between the renormalized Fermi constant, the weak mixing angle, and
the $W$-boson mass. The Fermi constant is measured most accurately in
muon decay. By considering $W$-boson mediation at scales well below
the $W$-boson mass, the bare effective Lagrangian is found to have the
form
\begin{equation}
\L_{\rm eff} =
{{G_\mu}_b\over\sqrt2}\:\bar\psi^{\nu_\mu}_b\gamma_\mu(1-\gamma_5)
\psi^\mu_b\:\bar\psi^e_b\gamma^\mu(1-\gamma_5)\psi^{\nu_e}_b~.
\end{equation}
As usual this is equal to
\begin{equation}
\L_{\rm eff} = {G_\mu+\delta G_\mu\over\sqrt2} {Z^{\nu_\mu}_L}^\half
{Z^\mu_L}^\half {Z^e_L}^\half {Z^{\nu_e}_L}^\half
\:\bar\psi^{\nu_\mu}\gamma_\mu(1-\gamma_5)\psi^\mu\:\bar\psi^e \gamma^\mu
(1-\gamma_5)\psi^{\nu_e}~.
\end{equation}
The proper vertex function will in general contain other current
bilinear products besides $(V-A)(V-A)$. However, in the SM and the
MSSM these terms are suppressed by the small electron and muon masses,
so they can be safely ignored. The vertex function contains the
following contributions:

\vspace{.8cm}
\hspace{1in}\psfig{figure=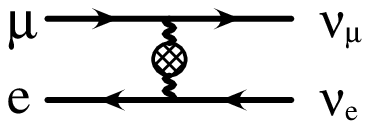,width=1.1in}

\vspace{-.42in}\hspace{2.15in}
$=~~i{G_\mu\over\sqrt2}~{\Pi^T_{WW}(0)\over M_W^2}$

\vspace{.3in}

\hspace{.5in}\psfig{figure=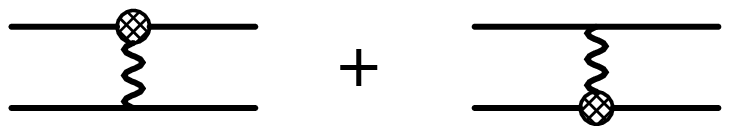,width=1.8in}

\vspace{-.37in}\hspace{2.54in}
$=~~i{G_\mu\over\sqrt2}~\delta_{\rm vertex}$

\vspace{.3in}

\hspace{.3in}\psfig{figure=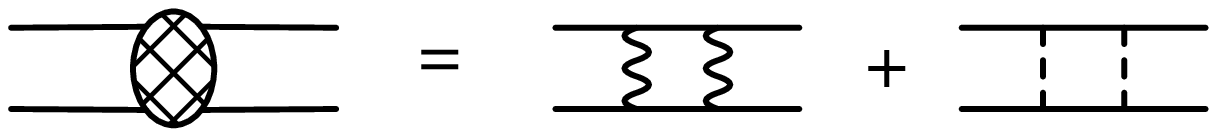,width=3in}

\vspace{-.35in}\hspace{3.45in}
$=~~i{G_\mu\over\sqrt2}~\delta_{\rm box}$

\vspace{.4in}

The renormalized coupling is then
\begin{equation}
{G_\mu\over\sqrt2}\biggl[1+{\delta G_\mu\over G_\mu} + \half\left(
\delta Z_L^\mu+\delta Z_L^e +\delta Z_L^{\nu_\mu}+\delta
Z_L^{\nu_e}\right) +{\Pi^T_{WW}(0)\over M_W^2} + \delta_{\rm vertex} +
\delta_{\rm box}\biggr]~.\label{Gmu}
\end{equation}
The muon decay rate was first
calculated in the effective QED theory, where the photon corrections
are taken into account, yielding
\begin{equation}
\tau^{-1}_\mu = {G_\mu^2m_\mu^5\over192\pi^3}\left(1-{8m_e^2\over
m_\mu^2}\right)\biggl[1-{\alpha\over2\pi}
\left({25\over4}-\pi^2\right)\biggr]~.\label{tau}
\end{equation}
The leading two-loop correction is included by evaluating $\alpha$ at
the scale $m_\mu$. It is customary to take Eq.~(\ref{tau}) as the
defining equation for $G_\mu$. Then, from (\ref{Gmu}), we determine
the counterterm $\delta G_\mu$,
\begin{eqnarray}
&&{G_\mu\over\sqrt2}\biggl[1+{\delta G_\mu\over G_\mu} + \half\left(
\delta Z_L^\mu+\delta Z_L^e +\delta Z_L^{\nu_\mu}+\delta
Z_L^{\nu_e}\right) + {\Pi^T_{WW}(0)\over M_W^2} + \delta_{\rm vertex}
+ \delta_{\rm box}\biggr]\nn\\
&&\qquad\qquad\qquad\qquad\qquad\qquad
= {G_\mu\over\sqrt2}\biggl[1+\delta_{\rm QED}\biggr]~.
\end{eqnarray}
$\delta_{\rm QED}$ includes the photon correction in the effective
theory,

\vspace{.2in}\hspace{1.4in}
\psfig{figure=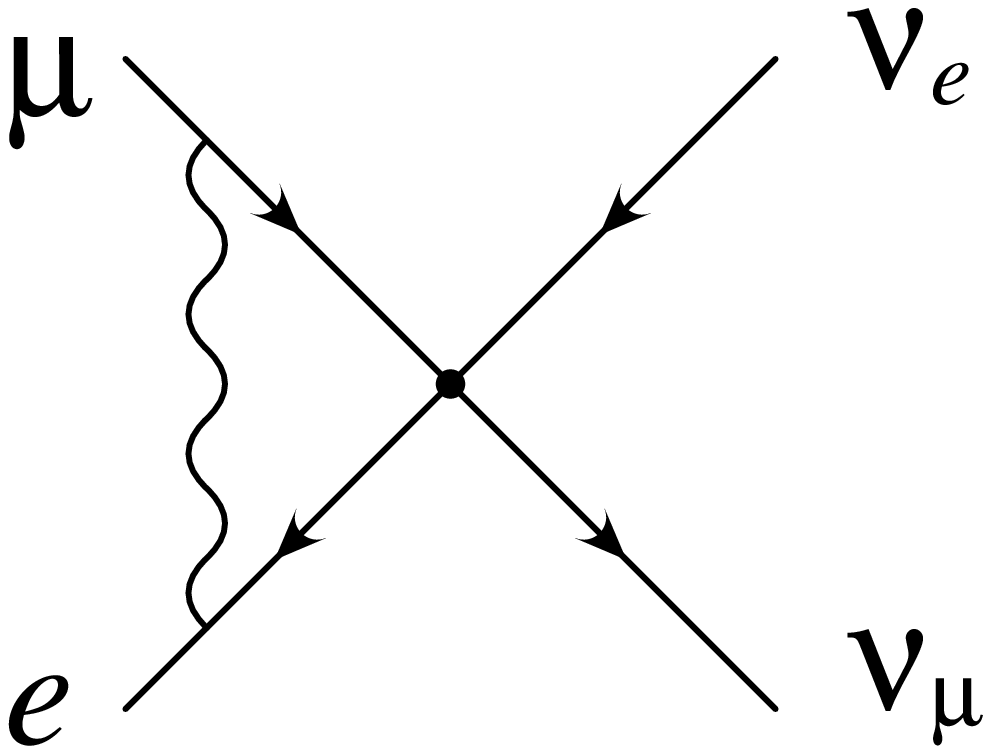,width=1.2in}

\vspace{-.65in}\hspace{2.45in}
$=~~i{G_\mu\over\sqrt2}~\delta_{\rm QED}$

\vspace{.50in}
\noindent Hence, we have
\begin{equation}
{\delta G_\mu\over G_\mu} = -{\Pi^T_{WW}(0)\over M_W^2} - \delta_{VB}
\label{dgmu}
\end{equation}
where $\delta_{VB}$ denotes the non-QED wave-function, vertex, and
box-diagram corrections,
\begin{equation}
\delta_{VB} = \half\left( \delta Z_L^\mu+\delta Z_L^e +\delta
Z_L^{\nu_\mu}+\delta Z_L^{\nu_e}\right) + \delta_{\rm vertex} +
\delta_{\rm box} - \delta_{\rm QED}~.
\end{equation}
The superpartner contributions to $\delta_{VB}$ are given in
Refs.~\citelow{GS,bmpz}.

We now have all the ingredients necessary to determine the
renormalized weak mixing angle.  The tree-level formula applies to the
bare parameters,
$$\sqrt2\,{G_\mu}_b={\pi\alpha_b\over {M_Z^2}_b s_b^2c_b^2}~.$$
Replacing the bare parameters by the renormalized ones plus the
counterterms, we find
\begin{equation}
\sqrt2\,(G_\mu+\delta G_\mu) = {\pi\,\alpha\,(1+\Delta\alpha)\over
(M_Z^2+\delta M_Z^2)(s^2+\delta s^2)(c^2+\delta c^2)}~.
\end{equation}
Expanding to first order, and using $\delta c^2=-\delta s^2$,
we have
\begin{equation}
s^2c^2\left[1+\delta s^2\left({c^2-s^2\over c^2s^2}\right)\right] =
{\pi\alpha\over\sqrt2\,G_\mu M_Z^2}\biggl[1+\Delta\alpha - {\delta
M_Z^2\over M_Z^2}-{\delta G_\mu\over G_\mu}\biggr]~.
\end{equation}
Because of the on-shell renormalization conditions we have applied,
the renormalized parameters on the right hand side of this equation
include the electric coupling measured in Thompson scattering, the
physical $Z$-boson pole mass measured at LEP, and the Fermi constant
determined from the measurement of the muon lifetime. All of these
physical quantities have corresponding counterterms. Plugging in the
counterterms from Eqs.~(\ref{dmz}), (\ref{de}) and (\ref{dgmu}), we
have
\begin{eqnarray}
&&s^2c^2\biggl[1+\delta s^2\left({c^2-s^2\over c^2s^2}\right)\biggr] 
=\label{s2c2}\\
&&{\pi\alpha\over\sqrt2\,G_\mu M_Z^2}\biggl[
1+\Pi^{T\prime}_{\gamma\gamma}(0) + 2{s\over c}{\Pi^T_{\gamma
Z}(0)\over M_Z^2} - \re{\Pi^T_{ZZ}(M_Z^2)\over M_Z^2} +
{\Pi^T_{WW}(0)\over M_W^2} + \delta_{VB}\biggr]~.\nn
\end{eqnarray}
We now specify a meaning for $s^2$ at one-loop level. This results in
a renormalization condition which determines the counterterm $\delta
s^2$.

All the counterterms we have encountered thus far have both ``finite''
and ``infinite'' parts. In the calculation of Feynman diagrams one
encounters divergences. A regularization method must be employed to
control these divergences. Dimensional reduction is a natural
regulator in supersymmetric theories, as will be discussed in the next
section. In dimensional reduction the integrals are made finite by
continuing spacetime to $n=4-2\epsilon$ dimensions. Then, the
divergences appear as poles in $1/\epsilon$. We can define a
renormalized quantity by requiring its counterterm to be purely
``infinite'', \ie\ proportional to $1/\epsilon$. The renormalized
quantity is referred to as a \dr\ parameter$\,$\footnote{The bar in
\dr~signifies that the $1/\epsilon$ pole is subtracted along with the
universal integral artifacts $\ln4\pi-\gamma_E$, where
$\gamma_E=0.577216...$ is Euler's constant.}, and depends on the \dr\
renormalization scale $Q$. From Eq.~(\ref{s2c2}) we find the \dr\
renormalized weak mixing angle,
\begin{equation}
\hat s^2(Q^2)\hat c^2(Q^2) =
{\pi\alpha\over\sqrt2\,G_\mu M_Z^2}\biggl[1+\Delta\hat\alpha -
\re{\hat\Pi^T_{ZZ}(M_Z^2)\over M_Z^2}+{\hat\Pi^T_{WW}(0)\over M_W^2} +
\hat\delta_{VB}\biggr]\label{s2c2f}~,
\end{equation}
where the hat signifies \dr\ renormalization. We define
\begin{equation}
\Delta\hat r_Z = \Delta\hat\alpha - \re{\hat\Pi^T_{ZZ}(M_Z^2)\over
M_Z^2}+{\hat\Pi^T_{WW}(0)\over M_W^2} + \hat\delta_{VB}~.
\end{equation}
The leading irreducible corrections are correctly resummed by
including the correction $\Delta \hat r_Z$ as \cite{dr resum}
\begin{equation}
\hat s^2\hat c^2 = {\pi\alpha\over\sqrt2\,G_\mu M_Z^2}
\biggl[{1\over1-\Delta\hat r_Z}\biggr]~.\label{sc hat}
\end{equation}
In particular, the largest correction is $\Delta\alpha$, and this form
corresponds to the solution of the one-loop electric charge
renormalization group equation.

To calculate the $W$-boson mass we introduce a \dr\ version of the
$\rho$ parameter,
\begin{equation}
\hat\rho = {1\over\hat c^2}{M_W^2\over M_Z^2} = {\hat M_Z^2\over\hat
M_W^2}{M_W^2\over M_Z^2} = 1+\Delta\hat\rho~.
\end{equation}
Using $\hat M_W^2 = M_W^2 + \re\hat\Pi_{WW}^T(M_W^2)$ and $\hat M_Z^2
= M_Z^2 + \re\hat\Pi_{ZZ}^T(M_Z^2)$, we find
\begin{equation}
\Delta\hat\rho = \re\biggl[{\hat\Pi^T_{ZZ}(M_Z^2)\over M_Z^2} -
{\hat\Pi^T_{WW}(M_W^2)\over M_W^2}\biggr]~.
\end{equation}
Defining 
\begin{equation}
c_0^2 = \half\left(1+\sqrt{1-{4\pi\alpha\over\sqrt2\,G_\mu M_Z^2}}\right)~,
\end{equation}
we can expand Eq.~(\ref{sc hat}) to obtain
\begin{equation}
\hat c^2 = c_0^2\biggl(1-{s_0^2\over c_0^2-s_0^2}\Delta \hat
r_Z\biggr)~.
\end{equation}
Then, using $M_W^2 = \hat c^2\hat\rho M_Z^2$, we find the prediction
for the $W$-boson pole mass,
\begin{equation}
M_W^2 = {c_0^2M_Z^2\over1-\Delta\hat\rho + {s_0^2\over
c_0^2-s_0^2}\Delta\hat r_Z}~.
\end{equation}
Since the $W$-boson mass is a physical quantity, the correction is
independent of the renormalization scheme to the order in which we are
working. In particular, $\Delta\rho + s_0^2/(s_0^2-c_0^2)\Delta r_Z$
is a physical observable. This combination of terms is gauge
invariant, renormalization scale independent, and is the same whether
hatted or unhatted.

\section{Regularization, threshold corrections, and the RGEs}

We have determined the mass, wave-function and electric charge
counterterms by applying the on-shell renormalization
conditions. These counterterms are comprised of two, three, and
four-point diagrams corresponding to the various physical processes
which are used to measure the input parameters. In supersymmetric
models with the standard model gauge symmetry (\eg\ the MSSM), the
various diagrams receive extra contributions from the superpartners,
but the forms of the counterterms remain unchanged.

The divergences which arise in calculating the counterterms and other
Feynman diagrams must be regulated. In order to preserve the Ward (or
Slavnov-Taylor) identities, and to avoid spurious complications, the
regularization method employed should preserve the symmetries of the
bare Lagrangian, even if they are spontaneously broken. Dimensional
regularization respects gauge symmetry, so it is widely used as a
regulator of gauge theories. In dimensional regularization spacetime
is continued to $n=4-2\epsilon$ dimensions. For $n<4$ it is clear that,
for example, the integral $$\int {d^n\!k\over k^4}$$ is ultraviolet
convergent. The divergences appear as simple poles in $\epsilon$.

In dimensional regularization \citeup{dreg} (known variously as `naive
dimensional reduction', DIMR, or DREG) the dimensionality of spacetime
and the (non-scalar) fields is continued to $n$-dimensions. Thus, the
index of the gauge field $A_\mu$ runs from $0$ to $n-1$. The Dirac
algebra is $n$-dimensional, so, for example, $\gamma_\mu\gamma^\mu=n$,
$\gamma_\mu\gamma_\nu\gamma^\mu=(2-n)\gamma_\nu$, and so on.

Continuing the dimensionality of the fields to $n$ dimensions poses
problems in supersymmetric theories. Supersymmetry requires equal
numbers of on-shell bosonic and fermionic degrees of freedom in a
multiplet. Continuing a four dimensional gauge supermultiplet to $n$
dimensions will spoil this equality, so naive dimensional
regularization (\eg\ \ms) breaks supersymmetry. One must add
counterterms order by order in perturbation theory to restore
supersymmetry. A new regularization method, called dimensional
reduction, was invented for supersymmetric theories. In dimensional
reduction \cite{dred} (DRED), spacetime is continued to $n$ dimensions
while the fields remain unchanged. Hence, the index on the gauge field
runs from 0 to 3, and the Dirac algebra is 4 dimensional. There are
potential problems with dimensional reduction \cite{drtj}. Ambiguities
appear in calculations using dimensional reduction at high orders in
perturbation theory. It is unknown whether these ambiguities show up
in calculations of corrections to physical observables. Even if they
did, there may exist a consistent prescription which gives correct and
unambiguous results. Also, it appears that the standard mass
factorization in QCD is violated in dimensional reduction
\cite{spira,beenakker}. The situation is unclear as there has not yet
been a complete investigation of the handling of collinear and
infrared singularities in massive QCD in dimensional reduction.  We
stress that for our purposes these potential problems are purely
academic. It is clear that, for the superpartner corrections we are
concerned with, dimensional reduction is a convenient and valid gauge
symmetry and supersymmetry preserving regulator at one or two loop
order.  We will use dimensional reduction in what follows.

We applied the on-shell renormalization conditions to determine the
mass, wave-function, charge and Fermi constant counterterms. As
mentioned in the previous section, it is in some cases expedient to
use minimal subtraction to define a renormalized quantity and its
counterterm. In minimal subtraction, naive dimensional regularization
is employed, and the counterterm is defined to be `purely infinite',
\ie\ proportional to the pole $1/\epsilon^n$. Usually the subscript
`MS' is used to signify a minimally subtracted renormalized
quantity. In modified minimal subtraction, the counterterm is
prescribed to contain the pole as well as the n-dimensional integral
artifacts that always appear with the pole, $\ln4\pi - \gamma_E$. The
two most common uses of modified minimal subtraction are in the
renormalization of the weak mixing angle and the strong coupling
constant. Such renormalized quantities are not physical
observables. They are model dependent and renormalization scale
dependent. Contributions from heavy particles do not decouple in such
quantities (one has to implement decoupling by hand). Nevertheless,
they are process independent, and they can be useful in comparing the
implications of different experimental measurements in a given model.

In supersymmetry we use dimensional reduction to regulate the
integrals, and the subtracted quantities in supersymmetric modified
minimal subtraction are referred to as \dr\ renormalized quantities.
We use a hat to denote a \dr\ renormalized quantity.

Consider the electric charge. If we perform \dr\ subtraction on the
bare charge we obtain the renormalized \dr\ charge,
\begin{eqnarray}
e_b = e + \delta e \qquad \Rightarrow \qquad e_b - (\delta e)_\infty &=&
\hat e(Q^2) = e + \delta e - (\delta e)_\infty \\
&=& e\left(1 +
\half\hat{\Pi}^{T\prime}_{\gamma\gamma}(0) + {s\over
c}{\hat\Pi^T_{\gamma Z}(0)\over M_Z^2}\right)~,\nn
\end{eqnarray}
where the $\infty$ subscript denotes the
$(1/\epsilon+\ln4\pi-\gamma_E)$ part.

\subsection{One-loop integrals}

We now discuss the calculation of Feynman diagrams.  With $n\le3$
there is only one scalar function for each $n$-point diagram which
needs to be determined to evaluate the one-loop Feynman integrals. All
the various tensor integrals can be written in terms of the scalar
functions.  The functions are as defined by Passarino and Veltman
\cite{PV}, except we work in the metric $(1,-1,-1,-1)$ and in some
cases we differ by a minus sign.

For the one-point diagram, we have
\begin{equation}
Q^{4-n}\int{d^n\!q\over(2\pi)^n}{1\over \left[q^2-m^2+i\varepsilon\right]}
\equiv{i\over16\pi^2}A_0(m^2)~.
\end{equation}
The renormalization scale $Q$ is necessary in order to keep the
$n$-dimensional coupling dimensionless. This integral is
\begin{equation}
A_0(m^2) = m^2\left({1\over\hat\epsilon} + 1 + \ln{Q^2\over
m^2}\right)~,
\end{equation}
where $1/\hat\epsilon = 1/\epsilon+\ln4\pi-\gamma_E$.  The scalar
two-point integral is defined as
\begin{equation}
Q^{4-n}\int{d^n\!q\over(2\pi)^n}{1\over\left[q^2-m_1^2+i\varepsilon\right]
\left[(q-p)^2-m_2^2+i\varepsilon\right]}
\equiv{i\over16\pi^2}B_0(p^2,m_1^2,m_2^2)~.
\end{equation}
This function can be written as
\begin{equation}
B_0(p^2,m_1^2,m_2^2) =
{1\over\hat\epsilon}-\int_0^1dx\ln{(1-x)\,m_1^2+x\,m_2^2-x(1-x)\,p^2\over
Q^2}~.\label{B0}
\end{equation}
Such Feynman parameter integral forms are especially useful in
analytically evaluating the Passarino-Veltman functions in special
cases, such as when one of its arguments is zero. The explicit formula
for $B_0$ in the general case can be found in Ref.~\citelow{bmpz}.

The vector two-point function is defined as
\begin{equation}
Q^{4-n}\int{d^n\!q\over(2\pi)^n}{q^\mu\over\left[q^2-m_1^2+i\varepsilon\right]
\left[(q-p)^2-m_2^2+i\varepsilon\right]}\equiv
{i\over16\pi^2}B_1(p^2,m_1^2,m_2^2)p^\mu~.
\end{equation}
$B_1$ has the Feynman parameter integral representation
\begin{equation}
B_1(p^2,m_1^2,m_2^2) = {1\over2\hat\epsilon}-\int_0^1dx\,x\,
\ln{(1-x)\,m_1^2+x\,m_2^2-x(1-x)\,p^2\over Q^2}~.\label{B1}
\end{equation}
The tensor two-point functions are defined as
\begin{eqnarray}
&&Q^{4-n}\int{d^n\!q\over(2\pi)^n}{q^\mu q^\nu\over
\left[q^2-m_1^2+i\varepsilon\right]\left[(q-p)^2-m_2^2+i\varepsilon\right]}
\\&&\qquad={i\over16\pi^2}\left[B_{21}(p^2,m_1^2,m_2^2)p^\mu p^\nu +
B_{22}(p^2,m_1^2,m_2^2)g^{\mu\nu}\right]~.\nn
\end{eqnarray}

The vector and tensor two-point functions can be written in terms of
the scalar functions $A_0$ and $B_0$ as
\begin{eqnarray}
&&B_1(p^2,m_1^2,m_2^2) \\
&&\qquad= {1\over2p^2}\Biggl\{A_0(m_2^2)-A_0(m_1^2) +
(p^2+m_1^2-m_2^2)B_0(p^2,m_1^2,m_2^2)\Biggr\}~,\nn\\
&&B_{21}(p^2,m_1^2,m_2^2) \\
&&\qquad= {1\over3p^2}\Biggl\{2A_0(m_2^2)-A_0(m_1^2)
+\left(p^2+{m_1^2-5m_2^2\over4}\right)B_0(p^2,m_1^2,m_2^2)\nn\\
&&\qquad-{m_2^2-m_1^2\over p^2}\left[A_0(m_2^2)-A_0(m_1^2)
+\left({3\over4}p^2+m_1^2-m_2^2\right)B_0(p^2,m_1^2,m_2^2)\right]
\nonumber\\
&&\qquad-{1\over2}\left(m_1^2+m_2^2-{1\over3}p^2\right)\Biggr\}~,
\nonumber\\
&&B_{22}(p^2,m_1^2,m_2^2) \\
&&\qquad= {1\over6}\Biggl\{{1\over2}\biggl(A_0(m_1^2) +
A_0(m_2^2)\biggr) +
\left(m_1^2+m_2^2-{1\over2}p^2\right)B_0(p^2,m_1^2,m_2^2)\nn\\ 
&&\qquad+\,{m_2^2-m_1^2\over2p^2}\biggl[A_0(m_2^2)-A_0(m_1^2) -
(m_2^2-m_1^2) B_0(p^2,m_1^2,m_2^2)\biggr]\nonumber\\
&&\qquad+\,m_1^2+m_2^2-{1\over3}p^2\Biggr\}~.\nonumber
\end{eqnarray}

The three-point scalar functions depend on the three external momenta,
and the masses of the three internal lines. Hence, the functions take
as arguments $(k^2,p_1^2,p_2^2,m_1^2,m_2^2,m_3^2)$, where $m_1$ is the
mass of the internal line between $k$ and $p_1$, and $m_2$ is the mass
of the line between $p_1$ and $p_2$. The scalar function is
\begin{eqnarray}
&&Q^{4-n}\int{d^n\!q\over(2\pi)^n}{1\over
\left[q^2-m_1^2+i\varepsilon\right]\left[(q-p_1)^2-m_2^2+i\varepsilon\right]
\left[(q-p_1-p_2)^2-m_3^2+i\varepsilon\right]}\nn\\
&&\qquad\qquad\qquad\equiv{i\over16\pi^2}C_0~.
\end{eqnarray}
The vector and tensor three-point function integrals are defined as
\begin{eqnarray}
&&Q^{4-n}\int{d^n\!q\over(2\pi)^n}{q^\mu\over
\left[q^2-m_1^2+i\varepsilon\right]\left[(q-p_1)^2-m_2^2+i\varepsilon\right]
\left[(q-p_1-p_2)^2-m_3^2+i\varepsilon\right]}\nn\\
&&\qquad\qquad\qquad
\equiv{i\over16\pi^2}\biggl[C_{11}\,p_1^\mu+C_{12}\,p_2^\mu\biggr]~,
\end{eqnarray}
and
\begin{eqnarray}
&&Q^{4-n}\int{d^n\!q\over(2\pi)^n}{q^\mu q^\nu\over
\left[q^2-m_1^2+i\varepsilon\right]\left[(q-p_1)^2-m_2^2
+i\varepsilon\right]
\left[(q-p_1-p_2)^2-m_3^2+i\varepsilon\right]}\nn\\
&&\qquad\equiv{i\over16\pi^2}\biggl[
  C_{21}\,p_1^\mu p_1^\nu + C_{22}\,p_2^\mu p_2^\nu
+ C_{23}\,\left\{p_1^\mu p_2^\nu + p_2^\mu p_1^\nu\right\}
+ C_{24}\,g^{\mu\nu}\biggr]~.
\end{eqnarray}
Analytic formulae for the $C$-functions can be found in
Ref.~\citelow{C0}.  The evaluation of these functions can involve
large cancellations, so much care should be exercised in implementing
algorithms. A {\tt FORTRAN} package, {\tt FF}, is available which
accurately evaluates these integrals, and more \cite{Oldenborgh}.

It is much more important to be familiar with the definitions and
general properties of these functions than to be concerned with the
general formula for them. For example, notice the general feature that
the residue of the $1/\hat\epsilon$ pole of each function is the same
as the coefficient of $\ln Q^2$. In the calculation of a physical
quantity (such as the $W$-boson mass) the $1/\hat\epsilon$ divergences
cancel out. This cancellation exactly corresponds to the cancellation
of the renormalization scale dependence.

It is useful to note the behavior of the Passarino-Veltman functions
in the limit that one of the arguments is large. Looking at $B_0$, for
example, we can see from Eq.~(\ref{B0}) that in the limit $M\gg m$, 
\begin{equation}
B_0(M^2,m^2,m^2) = {1\over\hat\epsilon} - \ln{M^2\over Q^2}
+~{\rm const}~+ {\cal O}\left({m^2\over M^2}\right).
\end{equation}
This form is true for any ordering of arguments $M$ and $m$.  $2B_1$
also has the same form.

\subsubsection{Example: SUSY-QCD corrections to the top quark mass}

Given these definitions it is easy to evaluate one-loop Feynman
diagrams. For example, consider the QCD correction to the top quark
mass. The color factor is
\begin{equation}
T^a_{ik}T^a_{kj} = C_2(R)\delta_{ij} = {N_c^2-1\over2N_c}\delta_{ij}
 = {4\over3}\delta_{ij}~,
\end{equation}
and the diagram is evaluated as
\begin{eqnarray}
i\Sigma(\slash p) &=& -\left({4g_3^2\over3}\right)Q^{4-n}\int 
{d^n\!q\over(2\pi)^n} {\gamma_\mu(\slash q +
m_t)\gamma^\mu\over\left[(q^2-m_t^2)+i\varepsilon\right]
\left[(q-p)^2+i\varepsilon\right]} \nn\\
&=&-{i\over16\pi^2}\left({4g_3^2\over3}\right)\left[-2\slash p
B_1(p^2,m_t^2,0) + 4m_tB_0(p^2,m_t^2,0)\right]~.
\end{eqnarray}
From Eq.~(\ref{dm}) we have $\delta m = \half\left[\Sigma_L(m^2)
+\Sigma_R(m^2)\right]m + \re\,\Sigma_S(m^2)$, so the gluon
contribution to $\delta m_t$ is
\begin{equation}
\delta m_t =
-{g_3^2\over6\pi^2}\left[2B_0(m_t^2,m_t^2,0)-B_1(m_t^2,m_t^2,0)\right]m_t~.
\end{equation}
From the forms in Eqs.~(\ref{B0}) and (\ref{B1}), we determine
\begin{eqnarray}
B_0(m^2,m^2,0) &=& {1\over\hat\epsilon} + \ln{Q^2\over m^2} + 2~,\\
B_1(m^2,m^2,0) &=& {1\over2\hat\epsilon} + \half\ln{Q^2\over m^2}
+ {3\over2}~.
\end{eqnarray}
Hence, the gluon contribution to the top quark mass in \dr\
renormalization is
\begin{equation}
\delta\hat m_t = -{\alpha_s\over3\pi}\left[3\ln
\left({Q^2\over m_t^2}\right) + 5\right]m_t~.
\end{equation}

The top quark mass is an input parameter, and as such its counterterm
is renormalization scale and renormalization scheme {\em
dependent}. The combination of counterterms and proper functions in
the prediction of a physical quantity is renormalization scale and
renormalization scheme {\em independent}.

As a second example, consider the squark/gluino contribution to the
top quark self-energy. If we ignore squark mixing, the left-handed top
squark contributes
\begin{eqnarray}
i\Sigma(\slash p) &=&
{8g_3^2\over3}Q^{4-n}\int{d^n\!q\over(2\pi)^n}{P_R(\slash q+m_{\tilde
g})P_L\over\left[q^2-m_{\tilde
g}^2+i\varepsilon\right]\left[(q-p)^2-m_{\tilde
t_L}^2+i\varepsilon\right]}\\
&=&{8g_3^2\over3}\left({i\over16\pi^2}\right) B_1(p^2,m_{\tilde
g}^2,m_{\tilde t_L}^2)\,\slash p\,P_L~,
\end{eqnarray}
so
\begin{equation}
\Sigma_L(p^2) = {g_3^2\over6\pi^2}B_1(p^2,m_{\tilde g}^2,m_{\tilde
t_L}^2)~.
\end{equation}
Including the right-handed top squark contribution, we find the
correction to $\delta m_t$,
\begin{equation}
\delta m_t = {\alpha_s\over3\pi}\re\left(B_1(m_t^2,m_{\tilde
g}^2,m_{\tilde t_L}^2) + B_1(m_t^2,m_{\tilde g}^2,m_{\tilde
t_R}^2)\right)m_t~.
\end{equation}
From Eq.~(\ref{B1}), we see that in the limit $m_{\tilde t}\gg
m_{\tilde g},\ m_t$, the function $B_1$ becomes
\begin{equation}
B_1(m_t^2,m_{\tilde g}^2,m_{\tilde t}^2) = {1\over2\hat\epsilon} +
\half\ln\left({Q^2\over m_{\tilde t}^2}\right) + {1\over4} + {\cal
O}\left({m_{\tilde g}^2,m_t^2\over m_{\tilde t}^2}\right)
\end{equation}
After \dr\ renormalization the full SUSY-QCD top quark mass correction
is then, in the limit $m_{\tilde t} = m_{\tilde t_L} = m_{\tilde t_R}
\gg m_t,\ m_{\tilde g}$,
\begin{equation}
\delta\hat m_t = {\alpha_s\over3\pi}\left[-3\ln\left({Q^2\over
m_t^2}\right) - 5 + \ln\left({Q^2\over m_{\tilde t}^2}\right) +
{1\over2} + {\cal O}\left({m_{\tilde g}^2,m_t^2\over m_{\tilde
t}^2}\right)\right]m_t~.\label{mtcor}
\end{equation}
The pole mass is related to the running \dr\ mass as
\begin{equation}
m_t = \hat m_t(Q) - \delta\hat m_t~.
\end{equation}

\subsection{Threshold corrections and the RGE}

We have just calculated the SUSY-QCD correction to the top quark
mass. The top quark mass is an input parameter in the standard
model. It is not a prediction. Hence, the correction is $Q$-dependent,
and in order to obtain a finite result in four dimensions one must
prescribe a renormalization procedure. We used \dr\ renormalization.

We see that the correction (\ref{mtcor}) involves logarithms. If a
logarithmic correction is large the validity of perturbation theory
may be threatened, as the expansion parameter in the perturbation
series becomes $(\alpha/\pi)\log(M/m)$.  Fortunately, these
logarithmic corrections can be resummed to all orders in perturbation
theory.  The renormalization group equation serves this purpose.

The renormalization group equation for the top quark mass can be
derived from Eq.~(\ref{mtcor}). Since the top quark pole mass is scale
independent, we have
\begin{equation}
{d\,m_t\over d\ln Q^2} = 0\quad\Longrightarrow\quad {d\,\hat m_t\over
d\ln Q^2} = {d\,\delta\hat m_t\over d\ln Q^2} = -
{1\over16\pi^2}\left({8\over3}\hat g_3^2(Q) + \cdots\right)\hat m_t(Q)
\end{equation}
where the dots indicate gauge and Yukawa coupling contributions from
the other (non-QCD) loops. The RGE involves the $Q$-dependent \dr\ top
quark mass and the $Q$-dependent \dr\ gauge and Yukawa
couplings. There are RG equations for these couplings as well, and
together they form a set of coupled differential equations. There is
no closed form solution to these equations, but it is easy to solve
them numerically.

We will now examine how the logarithmic and the non-logarithmic
corrections in Eq.~(\ref{mtcor}) relate to the RGE. For simplicity we
will set $m_{\tilde g}=m_t$. If $m_{\tilde t}\gg m_t$, there is no
single scale where the correction (\ref{mtcor}) is small. At the scale
$m_t$, the squark loop contribution is proportional to $\ln(m_{\tilde
t}/m_t)$. At the scale $Q=m_{\tilde t}$ the gluon loop
contribution involves the same large logarithm. The RGE resums the
large logarithms. Suppose we know the top quark pole mass, and we want
to determine the top quark running mass at scales above the top squark
mass. To resum the logs in Eq.~(\ref{mtcor}) we first apply the gluon
correction at the scale $m_t$. Notice there is no logarithm at this
scale, so
\begin{equation}
\hat m_t(m_t) = \left(1-{5\hat\alpha_s(m_t)\over3\pi}\right)m_t~.
\end{equation}
Now we solve the RGE to evaluate this running mass at the top squark
scale. The full RGE for the top quark mass involves all the gauge and
Yukawa couplings and must be solved numerically. Here we will neglect
all but the strong coupling so that we can solve it directly. In the
effective theory below the top squark mass scale, only the gluon
contributes to the $m_t$ RGE. Hence, from Eq.~(\ref{mtcor}), the
relevant RGE is
\begin{equation}
{d\,\hat m_t(t)\over dt} = -{\hat g_3^2(t)\over4\pi^2}\hat m_t(t)~,
\label{dmtgl}
\end{equation}
where $t=\ln Q^2$.  At one-loop order, the strong coupling RGE is
\begin{equation}
{d\,\hat g_3(t)\over dt} = {b_3\over16\pi^2}\hat g_3^3(t)~.
\end{equation}
In the regime below the squark mass scale, but
above the gluino mass scale, we have $b_3=5$. The solution to this
equation is
\begin{equation}
\hat g_3^{-2}(Q_2) = \hat g_3^{-2}(Q_1) -
{b_3\over8\pi^2}\ln\left({Q_2^2\over Q_1^2}\right)~.
\end{equation}
We can plug this into Eq.~(\ref{dmtgl}) to solve for $\hat m_t(t)$.
We find
\begin{equation}
\hat m_t(Q) = \hat m_t(m_t)
\left({\hat\alpha_s(m_t)\over\hat\alpha_s(Q)}\right)^{2\over b_3}~.
\end{equation}
This expression resums the logarithmic gluonic corrections to the top
quark mass. It is good at any scale between the top quark mass and the
top squark mass. To determine the running top quark mass above the
squark mass scale, we must include the top squark loop correction. To
avoid a large logarithm, which could spoil perturbation theory, we
apply the top squark correction at the top squark mass scale. At this
scale the logarithm in the top squark correction vanishes. The running
top quark mass just above the top squark mass scale is
\begin{equation}
\hat m_t(m_{\tilde t}) = \Biggl[
\left(1-{5\hat\alpha_s(m_t)\over3\pi}\right)\left(
{\hat\alpha_s(m_t)\over\hat\alpha_s(m_{\tilde t})}\right)^{2\over5}
+{\hat\alpha_s(m_{\tilde t})\over3\pi}\left({1\over2}
+{\cal O}\left({m_t^2\over m_{\tilde t}^2}\right)\right)\Biggr]m_t~.
\end{equation}
From this point we can run the top quark mass to higher scales by
solving the full MSSM RGE. Notice that above the top squark mass scale
both the top quark mass RGE and the strong coupling RGE change,
due to the squark contributions.

This example illustrates how to use the RGE to resum large
logarithms. It also demonstrates the interplay between the RGE and the
logarithmic and non-logarithmic parts of the threshold corrections.

\section{$Z$-pole precision data and calculation of the effective couplings}

We have renormalized the electroweak sector of the standard model.
Since the MSSM has the same gauge symmetry as the standard model, this
renormalization also applies in the MSSM.  Consider the \dr\ weak
mixing angle, for example.  We replace the bare parameters with the
renormalized ones plus a shift
\begin{equation}
s^2_bc^2_b = \frac{\pi\alpha_b}{\sqrt 2\,G_{\mu_b}M^2_{Z_b}}
\quad\Rightarrow\quad \hat s^2\hat c^2 =
\frac{\pi\alpha}{\sqrt2\,(G_\mu+\delta\hat
G_\mu)(M^2_Z+\delta\hat M^2_Z)(1-\Delta\hat\alpha)} 
\end{equation}
to obtain
\begin{eqnarray}
\hat s^2\hat c^2 &=& {\pi\alpha\over\sqrt2\,G_\mu M^2_Z}~
\left[{1\over1-\Delta\hat r_Z}\right]~,\\
\Delta\hat r_Z &=& \Delta\hat\alpha - {\delta\hat M^2_Z\over M^2_Z} 
- {\delta\hat G_\mu\over G_\mu} \\
&=& \hat{\Pi}^{T\prime}_{\gamma\gamma}(0)
+ 2\,{s\over c}~{\hat\Pi^T_{\gamma Z}(0)\over M^2_Z} -
\re\,{\hat\Pi^T_{ZZ}(M^2_Z)\over M^2_Z} +
{\hat\Pi^T_{WW}(0)\over M^2_W} + \hat\delta_{VB}~.\nonumber
\end{eqnarray}
The first 4 terms in the last expression for $\Delta\hat r_Z$ form the
oblique part of the correction, \ie\ the gauge boson self-energy
contribution.  The last part, $\hat\delta_{VB}$, is the non-oblique
part of the correction.  In the MSSM both the oblique and non-oblique
terms receive superpartner loop contributions.

Applying this renormalization leads to predictions for observables in
the MSSM which can be compared with experiment.  Such comparisons can
show which regions of supersymmetry parameter space are favored, and
show that certain regions are inconsistent with the data.

We already determined the prediction for the $W$-boson mass.  We shall
now derive expressions for other precision observables.  Most of the
precision data is due to measurements of $Z$-boson production and
decay.  We write the effective $Z$-boson--fermion--anti-fermion vertex
as $-i\gamma_\mu(g_v^\ef-g_a^\ef\gamma_5) =
-i(g_L^\ef\,P_L+g_R^\ef\,P_R)$.  The effective couplings depend on the
effective charge and the effective weak mixing angle as
\begin{equation}
\begin{array}{rcl} 
g_v^\ef &=& \frac{e^\ef}{2s^\ef c^\ef}
\ (T_3-2Q(s^\ef)^2) , \\[2ex] 
g_a^\ef &=& \frac{e^\ef}{2s^\ef c^\ef}\ T_3~,
\end{array}\qquad
\begin{array}{rcl} g_L^\ef &=& \frac{e^\ef}{s^\ef c^\ef}
\ (T_3-Q(s^\ef)^2) \ , \\[2ex]
g_R^\ef &=&-\frac{e^\ef s^\ef}{c^\ef}\ Q~.
\end{array}
\end{equation}
We will derive expressions for $(s^\ef)^2$ and $(e^\ef/s^\ef c^\ef)$.
First we introduce related observables.

The fermion left-right asymmetry is
\begin{equation}
A_f = \frac{(g^\ef_{L_f})^2-(g^\ef_{R_f})^2}{(g^\ef_{L_f})^2+(g^\ef_{R_f})^2} 
= \frac{2g_{v_f}^\ef g_{a_f}^\ef}{(g^\ef_{v_f})^2+(g^\ef_{a_f})^2} \ .
\end{equation}
Note that this asymmetry is independent of the electric charge.  Hence
a measurement of $A_f$ is just a measurement of the effective weak
mixing angle, $(s_f^\ef)^2 = \sin^2\theta_f^{\rm eff}$.  In principle,
each fermion has its own weak mixing angle in its coupling to the $Z$.
Each fermion also couples with its own strength, which is measured by
the partial widths, $\Gamma_f$.  The partial widths are proportional
to $(g^\ef_{v f})^2+(g^\ef_{a f})^2$.

The polarized electron beam allows SLC to measure the initial and
final state asymmetries independently.  The left-right asymmetry
\begin{equation}
A_{LR} = \frac{\sigma_L-\sigma_R}{\sigma_L+\sigma_R} = A_e
\end{equation}
is independent of the final state couplings.  The polarized
forward-backward asymmetry, on the other hand, is independent of the
initial state couplings,
\begin{equation}
A^{\rm pol}_{FB} = \frac{1}{P}\
\frac{(N_{P,F}-N_{-P,F})-(N_{P,B}-N_{-P,B})}
{(N_{P,F}-N_{-P,F})+(N_{P,B}-N_{-P,B})} = \frac{3}{4}\, A_f \ .
\end{equation}
$P$ is the polarization defined by
$P=(P_{e^-}-P_{e^+})/(1-P_{e^-}P_{e^+})$ and $N_{P,F}$ ($N_{P,B}$) is
the number of fermions produced in the forward (backward) direction.
SLC measures $A_f$ for $f=e,\mu,\tau,b,c$. SLC also measures the heavy
quark observables $R_{b,c}=\Gamma_{b,c}/\Gamma_{\rm had}$ and
$A^{FB}_{b,c}$.

At LEP the beams are unpolarized.  Hence, when the
forward-backward asymmetry is measured,
\begin{equation}
A^{FB}_f = \frac{\sigma_F-\sigma_B}{\sigma_F+\sigma_B} = \frac{3}{4}\,
A_eA_f \ ,
\end{equation}
a combination of the initial and final state effective weak mixing
angles is determined.  There are other means to measure $A_e$
independently at LEP. For example, a measurement of $A^{FB}_e$
determines $A_e$. Also, they can extract both $A_e$ and $A_\tau$ from
the $\tau$-polarization measurement,
\begin{equation}
P_\tau(\cos\theta) = -\frac{A_\tau+A_ef(\cos\theta)}{1+A_\tau
A_ef(\cos\theta)}\ , \qquad f(x) =
\frac{2x}{1+x^2} \ .
\end{equation}
Of course the various partial widths are also measured at LEP.  The
measurements include
\begin{equation}
A^{FB}_f,\; f = e,\mu,\tau,b,c~,\nn
\end{equation}
\begin{equation}
A_{e,\tau}~,\nn
\end{equation}
\begin{equation}
\Gamma_Z \propto \sum_f (g^\ef_{v_f})^2+(g^\ef_{a_f})^2~,\nn
\end{equation}
\begin{equation}
\sigma_{\rm had} = {12\pi\over M^2_Z}\ {\Gamma_e\Gamma_{\rm
had}\over\Gamma^2_Z}~,\nn
\end{equation}
\begin{equation}
R_\ell = \frac{\Gamma_{\rm had}}{\Gamma_\ell}
~, \quad \ell = e,\mu,\tau~,\nn
\end{equation}
\begin{equation}
R_{b,c} = \frac{\Gamma_{b,c}}{\Gamma_{\rm had}}~,\nn
\end{equation}
as well as $\sin^2\theta_e^\ef$ from the forward-backward charge
asymmetry $\left\langle Q_{FB}\right\rangle$.

The LEP and SLC collaborations do not present their results in a
model-independent fashion.  In extracting the values of observables
from the data they assume the gauge structure of the standard model.
That is, they assume the process $e^+e^-\rightarrow f\bar f$ is
mediated by standard model processes: $s$ and $t$ channel $\gamma$ and
$Z$ exchange, and box diagrams involving $\gamma$, $W$ and $Z$
exchange.  The diagrams involving the photon (the $s$- and $t$-channel
exchange, the photonic box diagrams, and the $\gamma-Z$ interference)
are taken into account by the experimentalists.  In other words, these
contributions are subtracted from the data.  The photonic corrections
are small.  On the $Z$-pole they are suppressed by a factor
$\Gamma_Z/M_Z$.  The genuine electroweak box diagrams are also
suppressed, and can be neglected.  In the SM and in the MSSM, the
remaining processes only involve the $Z$-fermion-antifermion
vertex. Thus, by subtracting the photon contributions the $Z$-vertices
are isolated.  In this way measurements of the coupling of the $Z$ to
$f\bar f$ are quoted.  It is good to keep in mind the model dependence
of the quoted results.  If other particles mediate the $e^+e^-
\rightarrow f\bar f$ process, the quoted values of the effective
$Z$-couplings are incorrect.  In the MSSM (with $R$-parity), however,
the only new processes which mediate $e^+e^-\rightarrow f\bar f$ are
superpartner box diagrams.  On the $Z$-pole these contributions are
negligible. Hence, the effective coupling analysis is ideally suited
to the MSSM. One needs only to compute the supersymmetric corrections
to the effective couplings, and compare to the quoted values.

Virtual loops involving superpartners change the effective $Zf\bar f$
couplings.  This is the only relevant supersymmetric correction in
$Z$-pole data.  We now derive the form of the $Z$-vertex
corrections. We apply the same renormalization procedure as in the
Sec.~\ref{sec.ren}.  The bare Lagrangian is of the form
\begin{equation}
\L = - \bar\psi_b\gamma_\mu(g_{v_b}-g_{a_b}\gamma_5)\, \psi_bZ^\mu_b -
e_bQ\, \bar\psi_b\gamma_\mu \psi_bA^\mu_b~.
\end{equation}
As usual, we apply wave-function renormalization and replace the bare
couplings by the renormalized couplings plus the counterterms,
\begin{eqnarray}
\L &=& -\bar\psi\gamma_\mu\biggl[(g_v+\delta g_v)
(Z_v-Z_a\gamma_5)-(g_a+\delta g_a)(Z_v-Z_a\gamma_5)\gamma_5\biggr]\psi\\
&&\times\biggl[Z^{\half}_{ZZ}Z^\mu+Z^{\half}_{Z\gamma}A^\mu\biggr]
-(e+\delta_e)Q\bar\psi\gamma_\mu(Z_v-Z_a\gamma_5)\psi
\biggl[Z^{\half}_{\gamma\gamma}A^\mu+ Z^{\half}_{\gamma
Z}Z^\mu\biggr]~.\nn
\end{eqnarray}
The $Z$-vertex may be read off, and at one-loop order we simply add
the proper three point function $\delta\Lambda(k^2,p_1^2,p_2^2)$ to
obtain the renormalized vertex (we neglect the magnetic moment
contribution)
\begin{eqnarray}
&&i\Gamma_\mu(k^2,p_1^2,p_2^2) =\label{renvert}\\
&& -i\gamma_\mu \Bigg[g_v\left(Z_v +
\frac{g_a}{g_v}\, Z_a +\frac{\delta g_v}{g_v}+\frac{1}{2}\ \delta
Z_{ZZ} + \frac{eQ}{g_v}\ Z^{\half}_{\gamma Z} +
\frac{\delta\Lambda_v(k^2,p_1^2,p_2^2)}{g_v}\right)\nn \\[2ex]
&&- g_a\left(Z_v + \frac{g_v}{g_a}\, Z_a + \frac{\delta g_a}{g_a} +
\frac{1}{2}\ \delta Z_{ZZ} +
\frac{\delta\Lambda_a(k^2,p_1^2,p_2^2)}{g_a}\right)\gamma_5\Bigg]~.\nn
\end{eqnarray}

At this point we can apply \dr\ renormalization to the couplings.  The
counterterms $\delta g_v$ and $\delta g_a$ are then purely `infinite'
(\ie\ proportional to $1/\hat\epsilon=1/\epsilon+\ln4\pi-\gamma_E$)
and are cancelled by the `infinite' parts of the other terms. The \dr\
couplings $\hat g_v$ and $\hat g_a$ are in tree-level relationship
with $\hat e$ and $\hat s$, which we have renormalized already. On the
$Z$-pole the effective axial coupling squared is
\begin{eqnarray}
(g^{\rm eff}_a)^2 &=& \frac{\hat e^2}{4\hat s^2\hat c^2} \ T^2_3\
\Biggl( 1-\left[\hat\Sigma_L(0)+\hat\Sigma_R(0)\right] - \frac{g_v}{g_a}
\,\left[\hat\Sigma_L(0)-\hat\Sigma_R(0)\right] \nonumber \\[2ex] 
&&\qquad\quad\quad\ - \re\, \hat{\Pi}^{T\prime}_{ZZ}(M^2_Z) + 2\,
\frac{\delta\hat\Lambda_a(M_Z^2,0,0)}{g_a}\Biggr)~,
\end{eqnarray}
where we have neglected the fermion mass.  From Eq.~(\ref{sc hat}) we
have
\begin{eqnarray}
\frac{\hat e^2}{4\hat s^2\hat c^2} &=& \sqrt2\ G_\mu M^2_Z
\left(1+\frac{\delta\hat M^2_Z}{M^2_Z}+\frac{\delta\hat
G_\mu}{G_\mu}\right) \nonumber \\[2ex]
&=& \sqrt2\, G_\mu M^2_Z \left(1+\re\,
\frac{\hat\Pi^T_{ZZ}(M^2_Z)}{M^2_Z} - \frac{\hat\Pi^T_{WW}(0)}{M^2_W}
- \hat\delta_{VB}\right)
\end{eqnarray}
so
\begin{eqnarray}
&&\hspace{-.12in}
(g^{\rm eff}_a)^2 = {\sqrt2\, G_\mu M^2_Z\over4}\ \Biggl(1+\re\,
{\hat\Pi^T_{ZZ}(M^2_Z)\over M^2_Z} -
\re\,\hat{\Pi}^{T\prime}_{ZZ}(M^2_Z) - {\hat\Pi^T_{WW}(0)\over M^2_W}\\[2ex] 
&&~~ - \left[\hat\Sigma_L(0) + \hat\Sigma_R(0)\right] - \frac{g_v}{g_a}\
\left[\hat\Sigma_L(0)-\hat\Sigma_R(0)\right] + 2
\ \frac{\delta\hat\Lambda_a(M_Z^2,0,0)}{g_a}- \hat\delta_{VB}\Biggr)~.\nn
\end{eqnarray}
$g_a^\ef$ determines the strength of the effective couplings.  The
correction to the effective mixing angle is determined by $g^{\rm
eff}_v/g^{\rm eff}_a$,
\begin{equation}
g^{\rm eff}_v = \frac{g^{\rm eff}_a}{T_3}\ (T_3-2Q\, s^2_{\rm eff})
\qquad \Rightarrow \qquad s^2_{\rm eff} = \frac{1}{4|Q|}\
\left(1-\frac{g^{\rm eff}_v}{g^{\rm eff}_a}\right) \ .
\end{equation}
We define $\hat \kappa$ to be the ratio of $s^2_{\rm eff}$ to $\hat
s^2$.  The correction to $\hat\kappa$ is given by
\begin{equation}
\Delta\hat\kappa = \left({1\over4|Q|^2s^2}-1\right) \
\left(\frac{\Delta\hat g_a}{g_a} - \frac{\Delta\hat g_v}{g_v}\right)~,
\end{equation}
where $\Delta\hat g_a/g_a$ and $\Delta\hat g_v/g_v$ are the
corrections inside the parentheses in Eq.~(\ref{renvert}), in the \dr\
scheme. Note the common $Z$ and vector fermion wave-function
renormalizations cancel out, leaving
\begin{eqnarray}
\Delta\hat\kappa &=&\left({1\over4|Q|^2s^2}-1\right)
\Bigg[{eQ\over g_v}\ \frac{\hat\Pi^T_{\gamma Z}(M^2_Z)}{M^2_Z}+
\frac{g^2_v-g^2_a}{2g_vg_a} \
\left[\hat\Sigma_L(0)-\hat\Sigma_R(0)\right]
\nn\\&&\qquad\qquad\qquad\qquad+\ \frac{\delta\hat\Lambda_a(M_Z^2,0,0)}{g_a} -
\frac{\delta\hat\Lambda_v(M_Z^2,0,0)}{g_v} \Bigg]\\[2ex] 
&=&\frac{c}{s}\ \frac{\hat\Pi^T_{\gamma Z}(M^2_Z)}{M^2_Z} +
\left({1\over4|Q|^2s^2}-1\right)\ \Bigg[ \frac{g^2_v-g^2_a}{2g_vg_a}\
\left[\hat\Sigma_L(0)-\hat\Sigma_R(0)\right]\nn\\
&&\qquad\qquad\qquad\qquad+\ \frac{\delta\hat\Lambda_a(M_Z^2,0,0)}{g_a} -
\frac{\delta\hat\Lambda_v(M_Z^2,0,0)}{g_v}\Bigg]~.\nn
\end{eqnarray}
We obtain the effective weak mixing angle by multiplying $\hat s^2$
from Eq.~(\ref{sc hat}) with $\hat\kappa = 1+\Delta\hat\kappa$.  The
most important supersymmetric contributions to $\delta\hat\Lambda_v$
and $\delta\hat\Lambda_a$ are listed for the $Zb\bar b$ coupling in
Ref.~\citelow{dlam}. Generalizing to the other fermion couplings is
straightforward.

\section{Comparison of supersymmetric models and precision data:
$\Delta\chi^2$ analysis}

The effective $Z$-boson couplings we just derived allow us to
calculate the following observables:
\begin{itemize}
\item{$Z$-width}
$$\Gamma_Z = \sum_f \Gamma_f = \sum_f
{\sqrt2\,G_\mu M_Z^3\over12\pi}\biggl[
(g_{vf}^{\rm eff})^2+(g_{af}^{\rm eff})^2\biggr]\biggl[
1+{\cal O}(\alpha)\biggr]$$
\item{Ratio of hadronic to leptonic width}
$$R_\ell = {\Gamma_{\rm had}\over\Gamma_\ell}$$
\item{Peak hadronic cross section}
$$\sigma_{\rm had} = {12\pi\over M_Z^2}
{\Gamma_e\Gamma_{\rm had}\over\Gamma_Z^2}$$
\item{Ratio of heavy quark to hadronic width}
$$R_{b,c} = {\Gamma_{b,c}\over\Gamma_{\rm had}}$$
\item{Left-right asymmetry}
$${\cal A}_f = {2g_{v\,f}^{\rm eff}g_{a\,f}^{\rm eff}\over
(g_{v\,f}^{\rm eff})^2+(g_{a\,f}^{\rm eff})^2}$$
\item{Forward-backward asymmetry}
$$A^{\rm FB}_f = {3\over4}{\cal A}_e{\cal A}_f$$
\item{Forward-backward left-right asymmetry}
$$A^{\rm FB}_{{\rm LR}\,f} = {\cal A}_f$$
\item{Effective weak mixing angle}
$$\sin^2\theta^{\rm eff}_f = {1\over4|Q_f|}
\left(1-{g^{\rm eff}_{v\,f}\over g^{\rm eff}_{a\,f}}\right)$$
\end{itemize}
The calculation of the effective $Z$-boson couplings allows
predictions for 20 precision observables. We can quantify the
comparison of a theory with the measurements by forming a $\chi^2$,
which gives a measure of the goodness of the fit.  We define the
$\chi^2$ below. The reliability of the results of this kind of test
increases with the number of observables. Hence, we will consider the
following 11 additional observables in the fit:

\begin{itemize}
\item{\bf 3 pole masses}

\noindent The $Z$-mass measurement is now as precise as the $G_\mu$
measurement. We fix $G_\mu$, but include $M_Z$ in the fit since its
error is correlated with other measurements.  The combined $W$-mass
measurements from CDF, D0, UA2, and LEP 2, give \cite{MW}
$$M_W = 80.430\pm0.076 {~\rm GeV}.$$
The combination of results of all top decay channels at CDF and D0
gives \cite{mt}
$$m_t = 175\pm5 {~\rm GeV}.$$
We take $M_Z$ and $m_t$ as inputs.

\item{\bf 6 low energy observables}

\noindent The weak charges of Tl and Cs have been measured in atomic
parity violation experiments (APV), giving \cite{apv} $Q_W(^{205}{\rm
Tl})=-114.77\pm1.23\pm3.44$, and $Q_W(^{133}{\rm Cs}) =
-72.11\pm0.27\pm0.89$. The deep inelastic scattering (DIS) experiments
have produced a measurement of a linear combination of effective
4-Fermi operator coefficients \cite{DIS}, $\kappa({\rm
DIS})=0.5805\pm0.0039$. The neutrino scattering experiments yield
determinations of leptonic 4-Fermi operator coefficients \cite{nue},
$g_V^{\nu e}=-0.041\pm0.015$ and $g_A^{\nu
e}=-0.507\pm0.014$. Finally, the CLEO measurement \cite{CLEO} of
B($B\rightarrow X_s\gamma$) yields the 90\% confidence level interval
$1\times 10^{-4} < {\rm B}(B\rightarrow X_s\gamma) < 4.2\times
10^{-4}$.  When an observable $x$ is defined in a finite interval,
there are arguments which suggest that a logistic transformation
$y={\rm lg}(x)$ should be performed, so that the new variable $y$ is
defined on the entire real axis. Including the transformed variable in
the $\chi^2$ results in a more Gaussian shaped distribution in that
variable. Hence, we include ${\rm lg}({\rm B}(B\rightarrow
X_s\gamma))$ in the $\chi^2$. For a variable defined on (0,1) the
logistic transformation is ${\rm lg}(x)=\ln(x/(1-x))$.

\item{\bf 2 gauge couplings}

\noindent We use the constraint $\Delta\alpha_{\rm had}^{(5)} =
0.02817 \pm 0.00062$ \cite{ad}. We also include the constraint
$\alpha_s(M_Z)=0.118\pm0.003$, which we obtain by combining non-$Z$
lineshape data \cite{as review}.
\end{itemize}
These 11 observables, combined with the 20 effective coupling
observables, give a total of 31 observables.

\subsection{Global fit in the standard model}

The measurements and experimental errors of the observables are listed
in Table~\ref{smfit}, along with the best fit values of the
observables in the standard model. Input data are as of August,
1997. The fit is performed as follows. We construct the $\chi^2$ by
adding the square of the deviations,
$$\chi^2 = \sum_{i=1}^{31} \left({o_i^{\rm meas}-o_i^{\rm pred}\over
\delta o_i} \right)^2~,$$ where $o_i^{\rm meas}$ and $\delta o_i$ is
the central value and error of the measurement of the $i$'th
observable, and $o_i^{\rm pred}$ is the prediction. Four of the
entries in Table~\ref{smfit} are inputs, so they do not have a
prediction associated with them. In the corresponding term in the
$\chi^2$ in place of the predicted value we use the input value.
\begin{table}[htb]
\begin{center}
\begin{tabular}{|l|c|c|r|} \hline
 & measurement & SM & pull \\ 
\hline \hline
$M_Z$ [GeV] &          $91.1867 \pm 0.0020$ & 91.1867 &    0.0 \\
$\Gamma_Z$ [GeV] &     $ 2.4948 \pm 0.0025$ &  2.4959 & $-0.4$ \\
$\sigma_{\rm had}$[nb]&$41.486  \pm 0.053 $ & 41.478  &    0.2 \\
$R_e$ &                $ 20.757 \pm 0.056 $ & 20.744  &    0.2 \\
$R_\mu$ &              $ 20.783 \pm 0.037 $ & 20.744  &    1.1 \\
$R_\tau$ &             $ 20.823 \pm 0.050 $ & 20.789  &    0.7 \\
$A^{FB} (e)$ &         $ 0.0160 \pm 0.0024$ &  0.0163 & $-0.1$ \\
$A^{FB} (\mu)$ &       $ 0.0163 \pm 0.0014$ &  0.0163 &    0.0 \\
$A^{FB} (\tau)$ &      $ 0.0192 \pm 0.0018$ &  0.0163 &    1.6 \\
\hline                      
${\cal P} (\tau)$ &    $ 0.1411 \pm 0.0064$ &  0.1476 & $-1.0$ \\
${\cal P}^{FB} (\tau)$&$ 0.1399 \pm 0.0073$ &  0.1476 & $-1.1$ \\
$\sin^2 \theta_{\rm eff}^e(Q^{FB})$ &
                       $ 0.2322 \pm 0.0010$ &  0.2315 &    0.8 \\
\hline
$R_b$ &                $ 0.2170 \pm 0.0009$ &  0.2158 &    1.3 \\
$R_c$ &                $ 0.1734 \pm 0.0048$ &  0.1722 &    0.2 \\
$A^{FB} (b)$ &         $ 0.0984 \pm 0.0024$ &  0.1035 & $-2.1$ \\
$A^{FB} (c)$ &         $ 0.0741 \pm 0.0048$ &  0.0739 &   0.0  \\
$A_{LR}^{FB} (b)$ &    $ 0.900  \pm 0.050 $ &  0.935  & $-0.7$ \\
$A_{LR}^{FB} (c)$ &    $ 0.650  \pm 0.058 $ &  0.668  & $-0.3$ \\
\hline
$A_e$ &                $ 0.1548 \pm 0.0033$ &  0.1476 &    2.2 \\
$A_{LR}^{FB} (\mu)$ &  $ 0.102  \pm 0.034 $ &  0.148  & $-1.3$ \\
$A_{LR}^{FB} (\tau)$ & $ 0.195  \pm 0.034 $ &  0.148  &    1.4 \\
\hline 
$M_W$ [GeV] &          $80.430  \pm 0.076 $ & 80.386  &    0.6 \\
$m_t$ [GeV] &          $ 175    \pm 5     $ & 172     &    0.6 \\
\hline
$Q_W ({\rm Cs})$ &     $-72.11  \pm 0.93  $ &$-73.11$ &    1.1 \\
$Q_W ({\rm Tl})$ &     $-114.8  \pm 3.7   $ &$-116.7$ &    0.5 \\
$\kappa ({\rm DIS})$ & $ 0.581  \pm 0.0039$ & 0.583   & $-0.7$ \\
$g_V^{\nu e}$ &        $-0.041  \pm 0.015 $ &$-0.0396$& $-0.1$ \\
$g_A^{\nu e}$ &        $-0.507  \pm 0.014 $ &$-0.5064$& $ 0.0$ \\
$\lg(B(B\rightarrow X_s\gamma))$ &
		       $ -8.49  \pm 0.45  $ & $-7.99$ & $-1.1$ \\  
\hline
$\Delta\alpha_{\rm had}^{(5)}$ &
                       $ 0.02817\pm 0.00062$& 0.02817 &    0.0 \\
$\alpha_s (M_Z)$ &     $ 0.118  \pm 0.003 $ & 0.1195  & $-0.5$ \\
\hline \hline
\end{tabular}
\end{center}
\caption{Results of a global fit to the standard model. For each
observable, we list the experimental result, the best fit result in
the SM, and the pull. The pull is the difference between the measured
value and the prediction, divided by the error.}
\label{smfit}
\end{table}

\begin{figure}[t]
\centerline{\psfig{figure=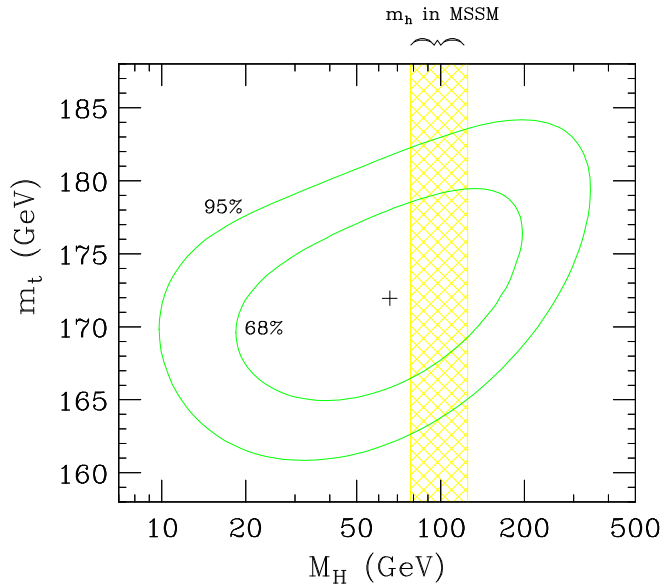,angle=90,width=3.4in}}
\caption{The 68 and 95\% C.L.\ contours in the ($m_t,~M_H$) plane in
the standard model. The range of Higgs boson mass predicted in the
MSSM is also shown.}
\label{fig:mh}
\end{figure}

We then minimize the $\chi^2$ function with respect to the inputs. In
the standard model, these are $M_Z$, $m_t$, $M_H$, $\Delta\alpha_{\rm
had}^{(5)}$, and $\alpha_s(M_Z)$. In the standard model the $\chi^2$
analysis gives rise to pertinent information about the Higgs boson
mass.  We find a standard model central value of $M_H=66$ GeV, and a
95\% confidence level upper bound of about 330 GeV. The 68\% and 95\%
confidence level contours are shown in Fig.~\ref{fig:mh} in the
($M_H,\ m_t$) plane.

In the MSSM, the lightest Higgs boson is constrained by the form of
the tree-level scalar potential to be below the $Z$-boson
mass. Radiative corrections raise the upper bound to about 125 GeV,
depending on the superpartner spectrum \cite{mh}. The lower bound
on $m_h$ in supersymmetric models is 78 GeV. We compare the range of
$m_h$ predicted in the MSSM with the best fit value in the SM in
Fig.~\ref{fig:mh}. We see that the range of values predicted in the
MSSM falls within the 68\% SM Higgs boson mass range.

\subsection{Global fit in supersymmetric models}

To discuss the global fit to the data in supersymmetry we have to
specify the soft supersymmetry breaking Lagrangian. Since we don't
know how supersymmetry is broken, we would, in the most general
analysis, consider the most general soft breaking terms. However, this
would lead to over one hundred parameters, with many complicated
correlations needed among the parameters to ensure compliance with
FCNC processes, rare decays, and so on. This is an intractable
approach. Instead, we will consider specific high scale models, in
which the soft Lagrangian takes a simple form. Besides reducing the
parameter space to a manageable set, these models have the virtue of
automatically satisfying the aforementioned constraints. We will
consider two types of models, which are distinguished by how the
supersymmetry breaking is communicated to the MSSM fields.

\subsubsection{Supersymmetric models}

In the canonical ``minimal supergravity model'' \cite{msugra},
supersymmetry breaking is parametrized by a universal $F$-term in a
hidden sector. This supersymmetry breaking vev is communicated to the
observable sector (\ie\ the MSSM fields) by gravitational effects,
giving rise to scalar masses, gaugino masses, and $A$-terms, all of
order $F/M_P\simeq100$--1000 GeV. Since gravity is flavor blind, it is
assumed that these mass parameters are universal.  Hence, in this
minimal model we take as inputs a universal scalar mass, $M_0$, a
universal gaugino mass, $M_{1/2}$, and a universal trilinear scalar
coupling, $A_0$. These inputs apply at the scale associated with the
messenger particle, in this case the Planck scale. However, for
simplicity, and to avoid model dependence, we will take these inputs
as valid at the scale where the U(1) and SU(2) gauge couplings unify,
$M_{\rm GUT}\simeq2\times10^{16}$ GeV.

In simple gauge-mediated models \cite{gmsb} supersymmetry is
dynamically broken in a hidden sector. Interactions between the
supersymmetry breaking sector and a standard model singlet, $S$, give
rise to vevs in the lowest and $F$ components of $S$. The singlet is
coupled to so-called messenger fields, $M$ and $\overline M$, through
a superpotential coupling $SM\overline M$. Through this coupling the
$S$ vevs produce diagonal (supersymmetry conserving) and off-diagonal
(supersymmetry breaking) entries in the $(M,\ \overline M)$ mass
matrix. The messenger fields are charged under the standard model
gauge symmetries. Hence, through the usual gauge interactions, the
supersymmetry breaking in the messenger fields is communicated to the
MSSM fields. This gives rise to gaugino and scalar masses. The masses
are proportional to the gauge couplings squared times the ratio of the
$F$ term of $S$ to its scalar component, $\Lambda=F/S$. The masses
also depend on the representation of the messenger fields. In order to
maintain the near unification of couplings in the MSSM, we will
consider complete SU(5) multiplets which do not affect gauge coupling
unification at one loop. We'll consider a messenger sector made of
$n_5$ \fiv\ fields and $n_{10}$ \ten\ fields. The gaugino masses and
the scalar masses squared are proportional to the effective number of
messenger \fiv\ fields, $n_5^{\rm eff} = n_5+3n_{10}$. The trilinear
$A$-terms are not generated at one-loop, so we set $A_0=0$ at the
messenger scale. We will consider models with $n_5^{\rm eff}=1$ (\ie\
$n_5=1,\ n_{10}=0$, the \fiv\ model) and $n_5^{\rm eff}=3$ (\ie\
$n_5=0,\ n_{10}=1$, the \ten\ model).

In both the gauge- and gravity-mediated models the spectrum of soft
masses is given at some high scale (either $M_{\rm GUT}$ or the
messenger scale $M$). In order to calculate the radiative corrections
to the precision observables we need to determine the physical
superpartner spectrum. We specify three steps to accomplish
this. First, we need to run the soft mass parameters from the initial
high scale down to the vicinity of the weak scale. As we describe
below, an appropriate scale to stop the running is the squark mass
scale. Second, we need to construct the weak scale mass matrices of
the charginos, neutralinos, squarks, sleptons and Higgs bosons. These
include the soft-breaking mass parameters and supersymmetry conserving
terms proportional to the Higgsino mass parameter $\mu$, the gauge
boson masses and/or the fermion masses.  We then determine the
physical superpartner mass spectrum and mixing angles by finding the
eigenvalues and eigenvectors of the mass matrices.

\subsection{Electroweak symmetry breaking}

In both the gravity- and gauge-mediated models we impose electroweak
symmetry breaking \cite{ewsb}. Electroweak symmetry breaking occurs
generically over the parameter space in both models. The RGEs of the
Higgs boson soft masses include terms proportional to the Yukawa
couplings. These terms drive the Higgs boson masses toward negative
values. If $\tan\beta$ is not very large, the top quark Yukawa
coupling is larger than the other Yukawa couplings, so the mass of the
Higgs boson which couples to the top quark ($H_2$) is driven
negative. This is exactly what is necessary to have electroweak
symmetry breaking take place if $\tan\beta>1$. In fact, because the
bottom Yukawa coupling can become larger than the top Yukawa coupling
at $\tan\beta\gsim60$, the mass of the Higgs which couples to the
bottom quark ($H_1$) can be driven too negative at large $\tan\beta$
and in this case electroweak symmetry breaking does not occur. This is
why large $\tan\beta$ values are excluded in our analysis.

Imposing electroweak symmetry breaking allows us to compute the heavy
Higgs boson mass and the Higgsino mass parameter $\mu$
\cite{bmpz}. Given the soft Higgs masses $m_{H_1}$ and $m_{H_2}$, we
can determine the CP-odd Higgs boson \dr\ mass $\hat m_A$, and
the \dr\ Higgsino mass parameter $\mu$,
\begin{eqnarray}
\hat m_A^2 &=& {\bar m_{H_2}^2-\bar m_{H_1}^2\over\cos2\beta}- \hat
M_Z^2~,\\ \mu^2 &=& {\tan^2\beta \bar m_{H_2}^2-\bar
m_{H_1}^2\over1-\tan^2\beta}- \half \hat M_Z^2~.
\end{eqnarray}
$\hat M_Z$ is the \dr\ $Z$-boson mass ($\hat M_Z^2 = M_Z^2 + \re
\hat\Pi_{ZZ}^T(M_Z^2)$), and $\bar m_{H_i} = m_{H_i} - t_i/v_i$. The
$t_i$ are the tadpole diagram contributions. These corrections are
necessary in order to ensure that we are at the minimum of the
one-loop effective potential. They are the effective potential
corrections in the diagrammatic approach. The contribution to
$t_i/v_i$ from a particle of mass $m$ which couples to $H_i$ is of the
form $m^2\log(m/Q)$, where $Q$ is the renormalization scale. Since the
squarks are typically the heaviest particles, and because of the color
and multiplicity factors, their contribution to the tadpole
corrections usually dominates. The corrections to electroweak symmetry
breaking can be applied most reliably at a scale $Q$ where the tadpole
corrections are minimized. This will generally be in the vicinity of
the squark mass scale.  This is why we stop the running of the soft
parameters at the squark mass scale. At $Q=m_{\tilde q}$ the
electroweak symmetry breaking conditions can be reliably computed. At
scales far from $m_{\tilde q}$, large logarithms make perturbation
theory less trustworthy.

We work in the convention that the gaugino masses are positive
and $\mu$ can be either sign, so the sign of $\mu$ needs to be
specified. To summarize, the parameter space of the minimal
supergravity model is
\begin{equation}
M_0,\ M_{1/2},\ A_0,\ \tan\beta,\ {\rm sgn}(\mu)~,\label{sug ps}
\end{equation}
and the gauge-mediated model parameter space is
\begin{equation}
M,\ \Lambda,\ \tan\beta,\ {\rm sgn}(\mu),\ n_5^{\rm eff}~.\label{gm ps}
\end{equation}

\subsection{Results}

We can now describe the results of the comparison of the predictions
of these supersymmetric models with the precision observables. A more
complete discussion can be found in Ref.~\citelow{EP}.  Choosing a
point in the parameter space of (\ref{sug ps}) or (\ref{gm ps}), we
run down the soft mass parameters to the squark mass scale, and check
for electroweak symmetry breaking. We then determine the physical
superpartner masses and mixings. We check if any of the superpartner
masses are below the current direct search limits. If so, we disregard
that point, and pick a new point. If all the mass bounds are
satisfied, we compute the supersymmetric correction to each
observable, and add it to the standard model prediction. We can then
evaluate $\chi^2$ and minimize it with respect to the standard model
input parameters, $M_Z$, $m_t$, $\Delta\alpha_{\rm had}^{(5)}$, and
$\alpha_s$. We have to iterate this process since the new values of
$M_Z$, $m_t$, $\Delta\alpha_{\rm had}^{(5)}$, and $\alpha_s$ after
minimization correspond to a different supersymmetric spectrum than
the one found in the previous iteration.

The supersymmetric corrections have the following characteristics. For
very large supersymmetric masses ($\tilde m\gg M_Z$) the
supersymmetric corrections to the weak-scale observables decouple, and
go to zero at least as fast as $M_Z/\tilde m$. In the large $\tilde m$
limit the predictions in any supersymmetric model will match those of
the standard model, with the standard model Higgs mass equal to the
light supersymmetric Higgs mass.  As the superpartner masses become
light, there can be relatively large corrections to the weak-scale
observables, and in general these corrections will upset the near
perfect standard model fit to the data shown in Table~\ref{smfit}.

\subsubsection{The oblique approximation}

Before looking at the full one-loop results, it is instructive to
consider an approximation. The full corrections can be divided up into
two sets of corrections, those from gauge boson self-energies, and the
vertex, wave-function and box diagram corrections. The first set, the
gauge-boson self-energy corrections, are called the oblique
corrections. They are universal in the sense that only certain
combinations of gauge boson self-energies appear in the corrections to
every physical observable. In fact, in the lowest order of a
derivative expansion, (where, for example, the derivative $\Pi'$ is
approximated by $[\Pi(M_Z^2)-\Pi(0)]/M_Z^2$) there are only three
combinations of gauge boson self-energies which appear. We consider
the combinations given by Peskin and Takeuchi \cite{PT}, the $S$, $T$
and $U$ parameters. They are given by
\begin{eqnarray}
S&=& \biggl[\cos^2\theta_W\left(F_{ZZ} - F_{\gamma\gamma}\right)
- {\cos\theta_W\over\sin\theta_W}\cos2\theta_W F_{\gamma Z}\biggr]
\times {4\sin^2\theta_W\over\alpha}\nonumber\\
T&=& \biggl[{\Pi_{WW}(0)\over M_W^2} - {\Pi_{ZZ}(0)\over M_Z^2}
-2{\sin\theta_W\over\cos\theta_W}{\Pi_{\gamma Z}(0)
\over M_Z^2}\biggr]\times {1\over\alpha}\nonumber\\
U&=&\biggl[F_{WW} - \cos^2\theta_WF_{ZZ} -
\sin^2\theta_WF_{\gamma\gamma}-\sin2\theta_WF_{\gamma Z}\biggr]
\times{4\sin^2\theta_W\over\alpha}\nonumber
\end{eqnarray}
where $F_{ij}=(\Pi_{ij}(M_j^2)-\Pi_{ij}(0))/M_j^2$ (except
$F_{\gamma\gamma}=\Pi_{\gamma\gamma}(M_Z^2)/M_Z^2$). In the oblique
approximation the non-oblique, process specific corrections (the
vertex corrections, wave function renormalization, and box diagrams)
are neglected. There is a simple argument why the oblique
approximation is expected to be a good approximation.  Every
superpartner (except the gluino) couples to the electroweak gauge
bosons, so every superpartner contributes to the oblique
corrections. Hence, the oblique corrections are enhanced by large
multiplicity factors. For example, there are 21 sfermion
contributions. In contrast, only certain superpartners contribute to
the non-oblique corrections. The non-oblique corrections are
constrained by the specific flavor of the fermions on the external
lines. Hence, the oblique corrections are expected in general to
dominate over the non-oblique corrections.

\begin{figure}[t]
\psfig{figure=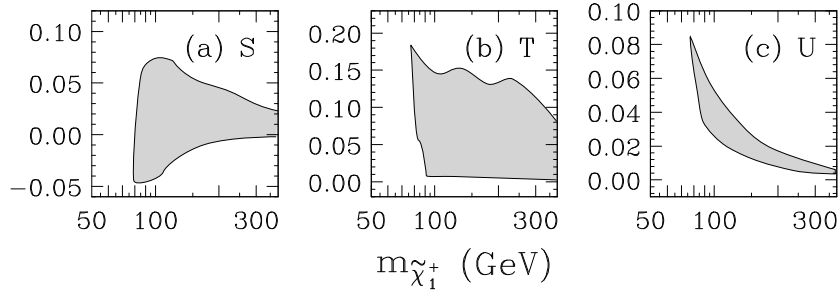,height=1.5in}
\caption{The range of the corrections to $S$, $T$ and $U$ in the
supergravity model, vs. the light chargino mass.}
\label{fig.STU}
\end{figure}

The oblique approximation gives an encapsulated view of the
supersymmetric corrections. Rather than looking at the corrections to
30 individual observables, one can simply compute the corrections to
the three oblique parameters, and then consider which observables are
sensitive to which oblique shifts. In Fig.~\ref{fig.STU} we show the
full range of the supersymmetric corrections (found in a random scan
of 50,000 points in parameters space) to the three oblique parameters
in the supergravity model, versus the light chargino mass. We see that
the corrections to $T$ and $U$ are always positive. The corrections to
$S$ and $U$ are small, less than 0.1, while the corrections to $T$ can
be almost as large as 0.2.

\begin{figure}[t]
\centerline{\psfig{figure=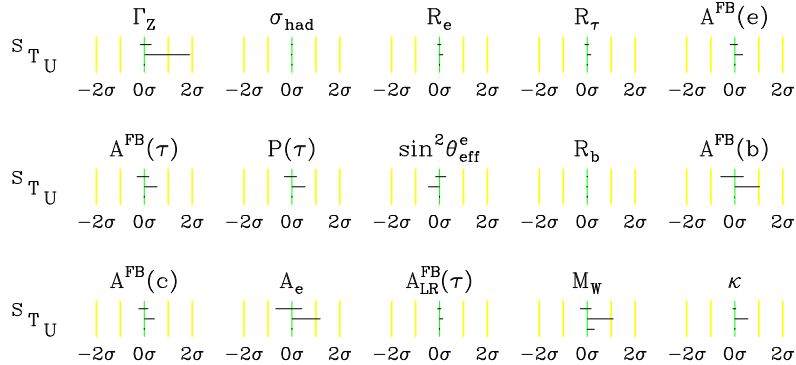,width=4.1in,angle=90}}
\caption{The range of the corrections to various observables due to
the oblique corrections to $S$, $T$ and $U$, in the supergravity
model. The corrections are divided by the experimental error.}
\label{fig.obl}
\end{figure}

These oblique shifts result in shifts in the prediction of each
observable. We can ascertain the importance of the oblique corrections
in the $\chi^2$ by dividing the correction to each observable (due to
the shift in each oblique parameter) by the experimental error. We
show the range of the oblique corrections to various observables in
the supergravity model in Fig.~\ref{fig.obl}. We see that some
observables, such as $\sigma_{\rm had}$, $R_e$ and $R_b$ receive
negligible oblique corrections, while others, such as $\Gamma_Z$,
$A_e$ and $M_W$, receive substantial oblique corrections$\,$\footnote{We
emphasize that the size of the corrections are measured against the
current experimental error. Fig.~\ref{fig.obl} does not reflect the
absolute size of the corrections.}.

\begin{figure}[t]
\centerline{\psfig{figure=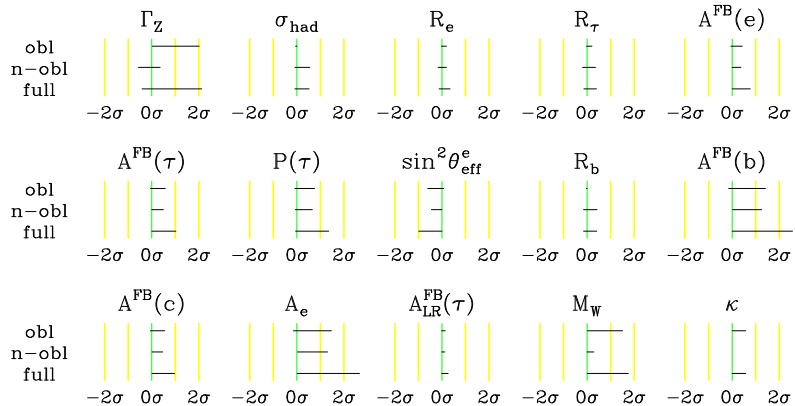,width=4.1in,angle=90}}
\caption{The range of the oblique ({\tt obl}), non-oblique ({\tt
n-obl}) and full corrections to various observables in the
supergravity model, divided by the experimental error. The non-oblique
correction to $\kappa$ is neglected.}
\label{fig.ono}
\end{figure}

There are two types of corrections in going from the oblique
approximation to the full one-loop corrections. Of course the
process-specific corrections must be added (the vertex, wave function,
and box diagram corrections). One also needs to correct for the full
gauge boson self-energies. As we have seen, some of the gauge boson
corrections enter as derivatives of the self-energies. In the oblique
approximation the derivatives are approximated by a difference of
self-energies. For the difference to be small, the particles in the
loop should be much heavier than the gauge bosons. Hence, the oblique
approximation should not work well when some superpartner masses are
light, of order $M_Z$. This is just what we find. In
Fig.~\ref{fig.ono} we compare the oblique, non-oblique and full
corrections in the supergravity model. We show the maximum range of
the corrections for various observables, in each case dividing by the
experimental error. We see that for most of the observables the
largest non-oblique corrections are about as large or larger than the
largest oblique corrections. The region of parameter space with light
superpartners has the largest corrections -- and it is just in this
region where the oblique approximation breaks down. Hence, when
excluding regions of parameter space it is crucial to include the full
one-loop corrections. We turn to the exclusion analysis next. We note
that the oblique corrections to the low-energy observables
$\kappa$(DIS), $g_V^{\nu e}$, $g_A^{\nu e}$, $Q_W(\rm Cs)$ and
$Q_W(\rm Tl)$ are all small relative to the experimental error. Hence,
the non-oblique corrections are bound to be small as well. We treat
these observables in the oblique approximation in the next
section. For all other observables the full one-loop corrections are
included.

\subsubsection{Excluded regions of supersymmetric parameter space}

We need to specify a criteria to exclude a point in supersymmetry
parameter space based on the $\chi^2$ test. We will use a criteria
analogous to that used in determining the standard model Higgs boson
mass limit. The 95\% confidence level limit on the Higgs boson mass
corresponds to $\Delta\chi^2=3.84$. That is, if the difference between
the minimum $\chi^2$ value and the $\chi^2$ value at a given Higgs
boson mass is greater than 3.84, that value of the Higgs boson mass is
excluded by the data with at least 95\% confidence. We adopt this
criteria directly to the supersymmetric models. In each model we
randomly scan over 50,000 points in the parameter space, and find the
minimum $\chi^2$, $\chi^2_{\rm min}$. Then, at each point we compute
$\Delta\chi^2 = \chi^2-\chi^2_{\rm min}$. If $\Delta\chi^2$ is larger
than 3.84 we consider this point in supersymmetry parameter space to
be excluded by the data. We emphasize that at each point we minimize
the $\chi^2$ with respect to $M_Z$, $m_t$, $\Delta\alpha_{\rm
had}^{(5)}$ and $\alpha_s$.

We show the results in two dimensional subspaces. In each plot the
dashed curve bounds the region of the parameter space which contains
points which are excluded. The solid line bounds the region of
parameter space containing points which are not excluded. Since we
project a multidimensional parameter space onto two dimensions, there
are regions which contain both excluded and unexcluded points in each
figure. The regions bounded by the dashed curves which are outside the
solid curves are absolutely excluded, independently of the values of
other parameters.  In the following we will focus on these absolutely
excluded regions.

\begin{figure}[t]
\psfig{figure=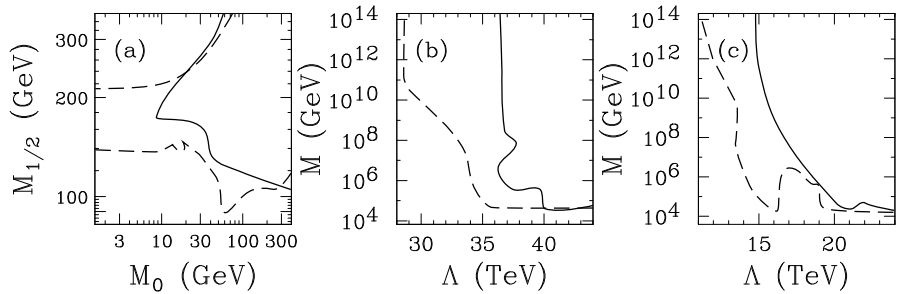,height=1.5in}
\caption{Excluded and non-excluded regions in the (a) supergravity
model, (b) \fiv\ gauge-mediated model, and (c) \ten\ gauge-mediated
model, in the ($M_{1/2}$, $M_0$) or ($M$, $\Lambda$) planes, with
$\mu>0$. The region of parameter space where it is possible to find
$\Delta\chi^2>3.84$ is bounded by the dahsed curve. The solid curve
indicates the region of parameter space within which it is possible to
find $\Delta\chi^2<3.84$. The region outside the solid curve, but
inside the dashed curve, is absolutely excluded, independently of the
values of any other parameters.}
\label{fig:ps}
\end{figure}

In each of the following figures, we show the results for the
supergravity model, the \fiv\ gauge-mediated model, and the \ten\
gauge-mediated model, in the three panes from left to right,
respectively.  In each figure we show the results for $\mu>0$. As
expected, the region of parameter space with the lightest
superpartners is excluded. This is exemplified by considering the
input parameters which directly set the scale of the superpartner
spectrum. In Fig.~\ref{fig:ps} we show the results in the $(M_{1/2},~
M_0)$ plane (Fig.~(a)), and in the $(M,~\Lambda)$ plane (Figs.~(b) and
(c)). We see that in the supergravity model we exclude $M_0<9$ GeV and
$M_{1/2}<105$ GeV for this sign of $\mu$. The region of parameter
space with the smallest values of $\Lambda$ is likewise excluded,
giving the limit $\Lambda>36 (14)$ TeV in the \fiv\ (\ten) model.

\begin{figure}[t]
\psfig{figure=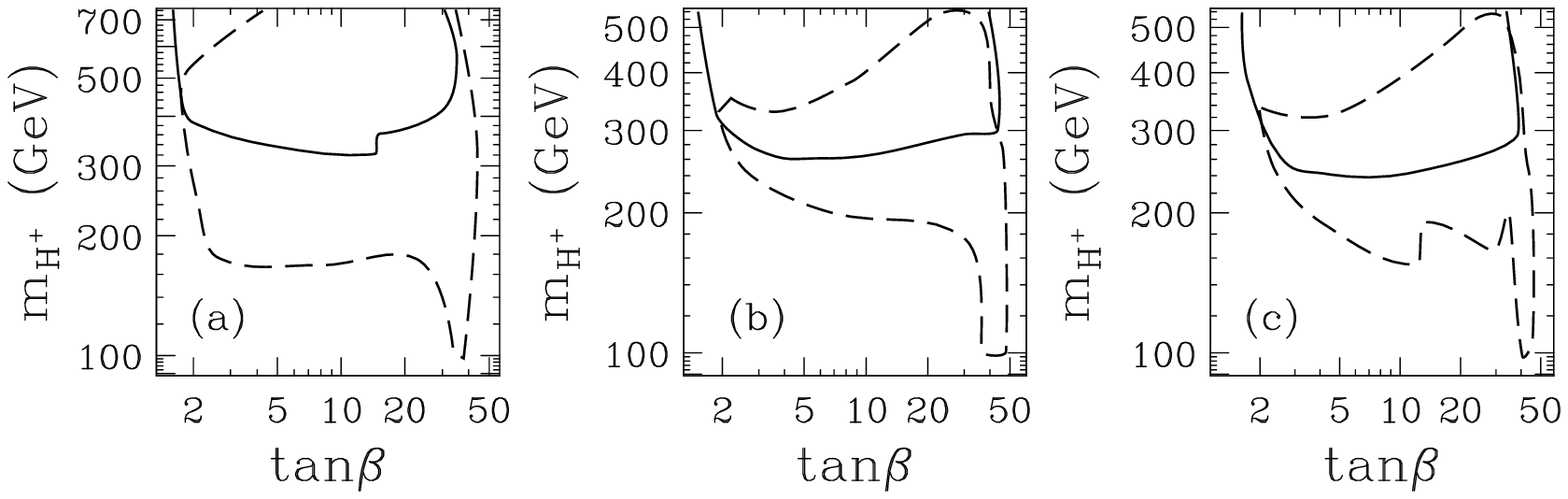,height=1.5in}
\caption{Excluded and non-excluded regions in the (charged Higgs mass,
$\tan\beta$) plane, in the (a) supergravity model, (b) \fiv\
gauge-mediated model, and (c) \ten\ gauge-mediated model. The curves
are as described in Fig.~\ref{fig:ps}.}
\label{fig:mhp}
\end{figure}

\begin{figure}[t]
\psfig{figure=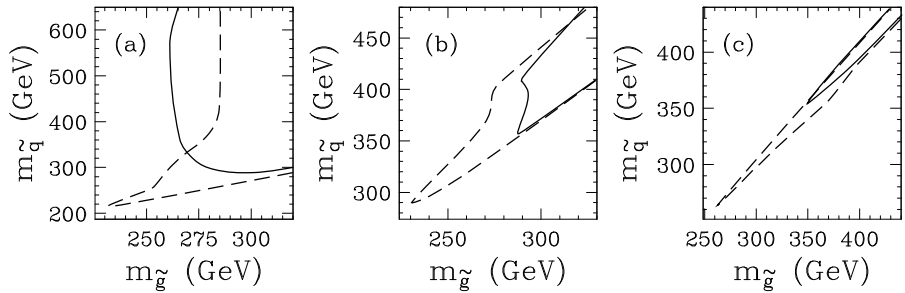,height=1.5in}
\caption{Excluded and non-excluded regions in the (squark mass, gluino
mass) plane, as in Fig.~\ref{fig:ps}.}
\label{fig:gq}
\end{figure}

These limits correspond to limits on the physical superpartner
spectrum. For example, in Fig.~\ref{fig:mhp} we show the results in
the $(m_{H^+},~\tan\beta)$ plane. The constraint from the
B$(B\rightarrow X_s\gamma)$ measurement strongly excludes light
charged Higgs masses in the $\mu>0$ case. Depending on the model,
charged Higgs masses below 240--330 GeV are excluded. As a last
example, we show the excluded region in the $(m_{\tilde q},~m_{\tilde
g})$ parameter space in Fig.~\ref{fig:gq}$\,$\footnote{The first two
generation squark masses are nearly degenerate. The squark mass
plotted is actually $m_{\tilde u_L}$.}. The squark (gluino) masses
must be above 280 GeV (255 GeV) in the supergravity model. In the
\ten\ gauge-mediated model the squark and gluino masses are excluded
below 340 GeV. See Ref.~\citelow{EP} for the excluded regions of
other masses and parameters.

The results in the $\mu<0$ case are typically not as strong, since the
B$(B\rightarrow X_s\gamma)$ constraint is much weaker. Nevertheless,
there are significant regions of parameter space excluded.  In fact,
for all three models under consideration we find the following bounds
on the superpartner spectrum for either sign of $\mu$:

\begin{itemize}
\item{$m_{\tilde q}>280$ GeV~,}
\item{$m_{\tilde g}>255$ GeV~,}
\item{$m_{\tilde\chi_1^0}>45$ GeV~,}
\item{$m_{\tilde\chi_2^+}>195$ GeV~,}
\item{$m_{\tilde e_L}>105$ GeV~,}
\item{$m_h>78$ GeV~,}
\item{$m_H>115$ GeV~,}
\item{$m_A>115$ GeV~,}
\item{$m_{H^+}>140$ GeV~.}
\end{itemize}

\noindent In each of these cases the $\chi^2$ analysis gives us
information about the particle spectrum beyond that which we can
currently obtain from direct particle searches. This provides an
example of an interesting and useful application of the
renormalization we derived in the previous section.

\section{Conclusions}

In these lectures we have reviewed the renormalization of the
electroweak sector of the standard model. This renormalization applies
directly to the MSSM. We reviewed the formalism which allows us to
compute the supersymmetric corrections to 21 LEP and SLC
observables. We discussed regularization, which is necessary to
calculate the Feynman diagrams. In particular we discussed \dr\
renormalization, which utilizes the dimensional reduction
regularization appropriate for supersymmetric theories. We then gave
some sample calculations, and illustrated the relationship between the
weak-scale threshold corrections and the renormalization group
equations.

Finally we compared the predictions for thirty-one observables in
three supersymmetric models with the data. We saw that the regions of
parameter space with light superpartners do not give a satisfactory
global fit to the data. We adopted an exclusion criteria which led to
lower bounds on the various superpartner masses and parameters. These
indirect limits are a nice complement to the limits found in the
ongoing direct production searches, and give us a better idea of
where to, and where not to, expect supersymmetry to show up.

\section*{Acknowledgments}
I thank the organizers Jon Bagger and K.T. Mahanthappa for all their
work which resulted in a very smoothly run and well organized
school. I acknowledge Jens Erler for his collaboration on the $\chi^2$
analysis.



\begin{thebibliography}{99}
\bibitem{note} 
Within recent years we have seen a 3.5 $\sigma$ deviation in $R_b$,
anomalous high $p_T$ jet production at Fermilab, anomalous 4-jet
events in ALEPH data and excess events at the highest energies in HERA
data. All are now consistent with the standard model.
\end{thebibliography}
\end{document}